\documentclass[aps, twocolumn, prl, showpacs,preprintnumbers,amsmath,amssymb,superscriptaddress]{revtex4-1}
\usepackage{amsmath,amsfonts,amssymb,graphics,graphicx,epsfig,color,times,indentfirst,layout,hyperref,subfigure,multirow,bm}
\usepackage{hyperref}
\usepackage{ulem}
\hypersetup{linktocpage,,colorlinks=true}
\usepackage{threeparttable}
\newcommand{\dg}{^{\dagger}}

\newcommand{\pmat}[1]{\begin{pmatrix}#1\end{pmatrix}}
\newcommand{\bk}{{\bf k}}
\newcommand{\bQ}{{\bf Q}}
\newcommand{\bl}{\hat {\bf l}}
\newcommand{\bM}{\hat {\bf m}}
\newcommand{\bn}{\hat {\bf n}}

\newcommand{\bx}{\hat {\bf x}}
\newcommand{\bw}{\hat {\mathbf \omega}}
\newcommand{\bp}{{\bf p}}
\newcommand{\bR}{\hat {\bf R}}

\newcommand{\bea}{\begin{eqnarray}}
\newcommand{\eea}{\end{eqnarray}}

\newcommand{\p}{\partial}

\newcommand{\nocontentsline}[3]{}
\newcommand{\tocless}[2]{\bgroup\let\addcontentsline=\nocontentsline#1{#2}\egroup}

\begin{document}

\title{Skyrme insulators: insulators at the brink of superconductivity}

\author{Onur Erten}
\affiliation{Center for Materials Theory, Rutgers University,
Piscataway, New Jersey, 08854, USA }
\affiliation{Max Planck Institute for the Physics of Complex Systems, D-01187 Dresden, Germany}

\author{Po-Yao Chang} 
\affiliation{Center for Materials Theory, Rutgers University, Piscataway, New Jersey, 08854, USA }

\author{Piers Coleman}
\affiliation{Center for Materials Theory, Rutgers University, Piscataway, New Jersey, 08854, USA }
\affiliation{Department of Physics, Royal Holloway, University of London, Egham, Surrey TW20 0EX, UK }

\author{Alexei M. Tsvelik}
\affiliation{Division of Condensed Matter Physics and Material Science, Brookhaven National Laboratory, Upton, NY 11973}

\begin{abstract}

Current theories of superfluidity are based on the idea of a coherent
quantum state with topologically protected, quantized
circulation. When this topological
protection is absent, as in the case of  $^{3}$He-A, the coherent
quantum state no longer supports persistent superflow. Here we argue
that the loss of topological protection in a superconductor gives rise
to an insulating ground state.  We specifically introduce the concept
of a {\it Skyrme insulator} to describe the coherent dielectric state 
that results from the topological failure of superflow carried by 
a complex vector order parameter.  
We apply this idea to the case of 
SmB$_{6}$, arguing that the observation of a diamagnetic Fermi
surface within an insulating bulk can be understood in terms of 
a Skyrme insulator. Our theory enables us to understand the linear
specific heat of SmB$_{6}$ in terms of a neutral Majorana Fermi sea
and leads us to predict that in low fields of order a Gauss, SmB$_{6}$
will develop a Meissner effect.
\end{abstract}

\maketitle
\begin{figure}[t]
\centering
\includegraphics[width= 6.5 cm] {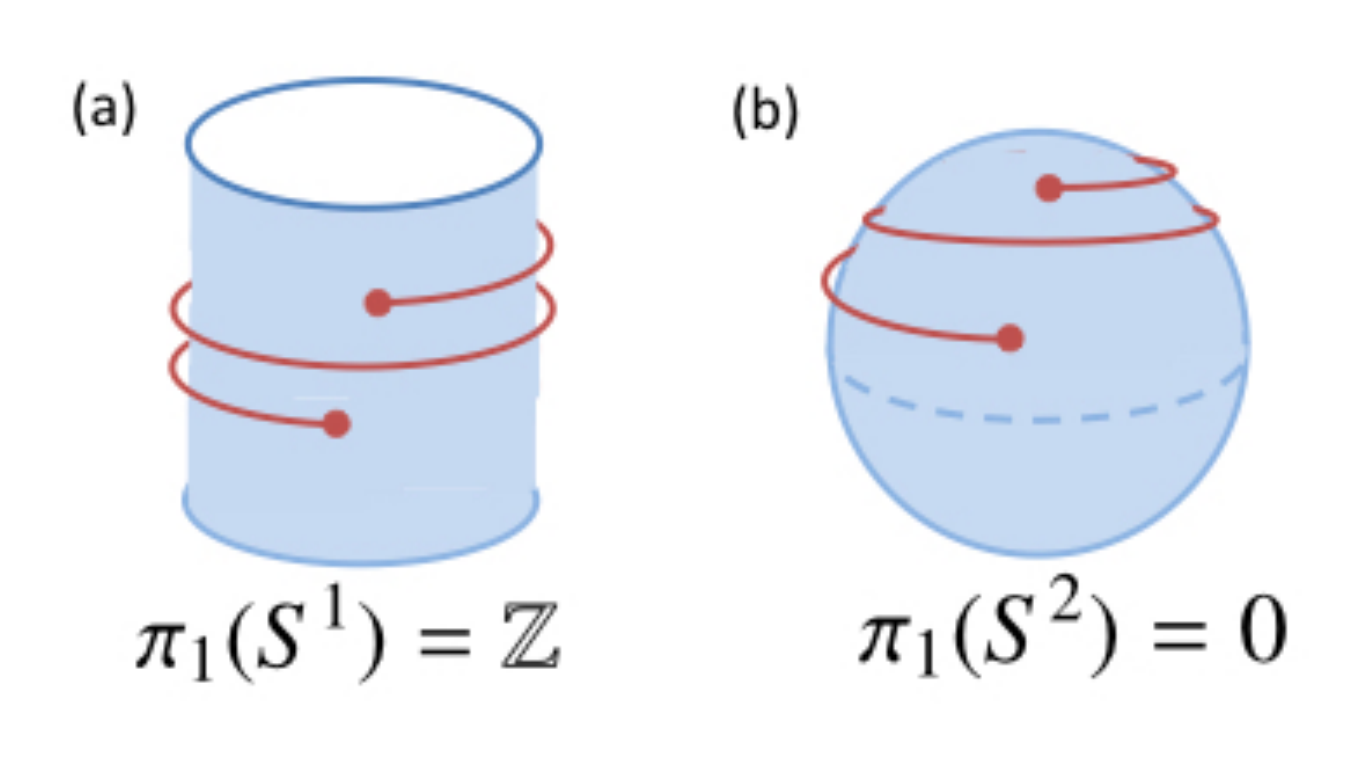}
\caption{Illustration of topological stability.  The stability of a
supercurrent is analogous to topological stability of a string wrapped
around a surface. (a) The winding number
of a string wrapped around a rod is topologically stable and it can not
be unravelled (b) A string wrapped around the equator of a sphere
unravels due to a lack of topological stability. 
}
\label{Fig1}
\end{figure}

While it is widely understood that superfluids and superconductors 
carry persistent ``supercurrents'' associated with the rigidity of the
broken symmetry condensate\cite{London37}, 
it is less commonly
appreciated that the remarkable persistence of supercurrents has its origins in 
topology. The order parameter of a 
conventional superfluid or superconductor
lies on a circular manifold ($S^{1}$), and the topologically stable
winding number of the order parameter, like a string wrapped multiple
times around a rod, protects a circulating superflow.  
However, if the order parameter lies on a
higher dimensional manifold, such as the surface of a sphere
($S^{2}$), then the winding
has no topological protection and 
putative supercurrents relax their  energy through a continuous
reduction of the winding number, leading to dissipation [see Fig. \ref{Fig1}]. This 
topological failure of superfluidity is observed in the A phase of
$^3$He, which exhibits dissipation \cite{Anderson:1977de,Bhattacharyya_PRL1997,Vollhardt_book,Volovik_book}.
Similar behavior has also been observed in spinor
Bose gases, where the decay of Rabi oscillations between two
condensates reveals the unravelling superflow\cite{Matthews_PRL1999}.

Here we propose an extension of this concept to superconductors, 
arguing that when a charge condensate fails 
to support a topologically stable circulation, the resulting 
medium forms a novel dielectric. 
Though our arguments enjoy general application,  they are specifically
motivated by the Kondo insulator,
SmB$_6$. While transport \cite{Wolgast_PRB2013, Kim_SciRep2013,
Kim_NatMat2014} and photoemission
\cite{Jiang_NatComm2013,Neupane_NatComm2013,
Xu_PRB2013, Frantzeskakis_PRB2013, Xi_NatComm2014} measurements
demonstrate that SmB$_{6}$ is an insulator with robust, likely 
topological surface states, the observation of bulk quantum oscillations \cite{Li_Science2014,
Sebastian_Science2015}, linear specific heat, anomalous thermal and ac optical
conductivity\cite{Flachbart_1982, Xu_arXiv2016,Sebastian_MarchMeeting2016,Laurita_PRB2016} 
have raised the fascinating possibility of a ``neutral'' Fermi surface
in the bulk, which nonetheless exhibits Landau quantization.
Landau quantization and the de-Haas van Alphen effect are normally understood
as a semi-classical quantization of cyclotron motion\cite{Onsager52}.
However, rather general arguments tell us that gauge invariance 
makes the Coulomb and Lorentz forces inseparable, 
so that quasiparticles that develop a Landau quantization must also respond to
an electric field, forming a metal.   To see this note 
that gauge invariance obliges particles to interact 
with the vector potential, entering into the gauge-invariant 
kinetic momentum $\pi = ({\bf p}- e{\bf A})$; the corresponding equation of
motion $d\pi/dt =q ( {\bf E + v\times  B})$ necessarily 
contains both $\bf E$ and $\bf B$ as temporal and spatial
gradients of the underlying vector potential. 
In other words unless the bulk somehow breaks gauge invariance, 
quantized cyclotron motion is incompatible 
with insulating behavior.  This robust line of reasoning motivates the
hypothesis that SmB$_{6}$ is a kind of failed superconductor, 
formed from a topological break-down of an underlying condensate. 
This paper examine the consequences of this line of reasoning, using
largely macroscopic arguments to make predictions that can be used
test this new hypothesis. 

General arguments tell us that the condition for the stability of a 
superfluid is determined by the order parameter manifold or ``coset
space'' $G/H$ formed between the symmetry group $G$ of the
Hamiltonian and the invariant subgroup $H$ of the order parameter.
The absence of coherent bulk superflow requires that the first homotopy class
$\pi_{1} (G/H)\neq {\mathbb{Z}}$ is sparse, lacking
the infinite set of integers which protect macroscopic winding of the phase.
This means that $G/H$ is a higher dimensional
non-Abelian coset space, most naturally formed through the
condensation of bosons or Cooper pairs with angular momentum.
Thus in spinor Bose gases, an atomic spinor condensate
lives on an $SU (2)$ manifold with 
$\pi_{1} (SU (2))=0$: in this case the observed decay of vorticity gives rise
to Rabi oscillations\cite{Matthews_PRL1999}.  Similarly, in 
superfluid $^{3}$He-A, an $SO (3)$ manifold associated with a
dipole-locked triplet paired state\cite{Vollhardt_book,Volovik_book},
for which $\pi_{1} (SO (3))=Z_{2}$ allows a single vortex, but no
macroscopic circulation in the bulk 

In the solid state, 
the conditions for a topological failure of superconductivity are
complicated by crystal anisotropy. 
On the one hand, if the 
condensate carries orbital angular momentum, it
will tend to lock to the lattice, 
collapsing the manifold back to $U (1)$. On the other hand,
if the order parameter has s-wave symmetry, its $U (1)$ coset space 
allows stable vortices. 

There are however
two ways around this no-go argument. 
The first, is if there is an additional
``isospin'' symmetry of the order parameter. For example, 
the half-filled attractive Hubbard model\cite{negativeU}, which 
forms a ``supersolid'' ground-state with 
a perfect spherical ($S^{2}$) manifold of degenerate
charge density and superconducting states, with pure superconductivity
along the equator and a pure density wave at the pole. 
In this special case, 
supercurrents can always decay into a density wave. 

A second route 
is suggested by crystal field theory, which allows the
restoration of crystalline isotropy for low spin objects, such as a spin 
1/2 ferromagnet in a cubic crystal. 
Were an analogous s-wave spin-triplet condensate to form,
isotropy would be assured. 
Rather general arguments 
suggest that the way to achieve an s-wave spin triplet, is through
the development of odd-frequency pairing. 
The Gorkov function of a triplet  condensate
has the  form
\begin{equation}\label{}
 {\bf d} (1-2)=\langle \psi_{\alpha } (1) (i\sigma_{2}\vec{\sigma })_{\alpha \beta }\psi_{\beta } (2)\rangle .
\end{equation}
where $1\equiv (\vec{x}_{1},t_{1})$ 
and $2\equiv (\vec{x}_{2},t_{2})$ 
are the space-time co-ordinates of the electrons.
Exchange statistics enforce the pair wavefunction ${\bf d} (X)=
- {\bf d} (-X)$ to be  odd under particle exchange. 
Conventionally,  ${\bf d} (\vec{x},t) = -{\bf d} (-\vec{x},t)
$ is an 
odd function of {\sl position}, leading to 
odd-angular momentum pairs. 
By contrast, an {\sl s-wave} triplet is  even in space and must
therefore be odd in time,
${\bf d} (|x|,t) = - {\bf d} (|x|,-t)$, as first proposed by Berezinsky
\cite{oddfreq1,oddfreq2,Coleman94,oddfreq3,efetov2005,eschrig}.  
Odd-frequency triplet pairing has been experimentally-established 
as a proximity effect in hybrid superconductor-ferromagnetic tunnel junctions\cite{efetov2005,eschrig}. 
But for spontaneous odd-frequency pairing, 
we need to identify an equal-time 
order parameter. Following \cite{oddfreq3},
we can do this by writing the time derivative of the Gorkov function using the
Heisenberg equation of motion: 
\begin{equation}\label{}
{\mathbf{\Psi }} (1) =\left.\frac{\partial{\bf d}
(1-2)}{\partial t_{1}}\right|_{1=2} = 
\langle [\psi_{\alpha } (1),H](\sigma_2\vec\sigma)_{\alpha  \beta }
\psi_{\beta } (1)\rangle .
\end{equation}
The specific form of this  composite operator  depends on the
microscopic physics, but the important point to notice is that it is
an equal-time expectation value which defines a complex
vector order parameter 
 ${\mathbf{ \Psi }}={\mathbf\Psi}_{1}+ i {\mathbf\Psi}_{2}$.

The case of SmB$_{6}$ motivates us to examine a  concrete example of this
idea. 
We consider a Kondo lattice of 
local moments (${\bf S}_{j}$) interacting with electrons via 
an exchange interaction of form $H=J \sum_{j}{\bf S}_{j}\cdot \psi\dg 
(x_{j})\vec{\sigma }\psi (x_{j})$. In this case, 
the crucial commutator has the form $[\psi_{\alpha } (x),H]=
J ({\bf S} (x)\cdot\vec{\sigma})_{\alpha \gamma}\psi_{\gamma} (x)$, giving rise to 
an equal-time, composite pair order parameter
between local moments and s-wave pairs \cite{Emery:1992ef,oddfreq3}
\begin{equation}\label{}
 {\mathbf \Psi} (x) \propto \langle \psi_{\uparrow} (x) \psi_{\downarrow}(x) {\bf S} (x)\rangle.
\end{equation}
In microscopic theory, it is actually more natural to consider
an antiferromagnetic version of composite order, formed between
the staggered magnetization and the
pair density,  $ {\mathbf \Psi} (x) = (-1)^{i+j+k} \langle \psi_{\uparrow} (x) \psi_{\downarrow}(x) {\bf S} (x)\rangle
$ 
\cite{
Emery:1992ef,Coleman94,zachar2001,Berg_PRL2010}. These details
do not however affect the development of the phenomenology. 

We now consider a general 
Ginzburg Landau free energy for an s-wave triplet condensate. 
Unlike a p-wave triplet, 
the absence of orbital components to the order parameter considerably
simpifies the Ginzburg Landau free energy density\cite{Knigavko_PRB1999},
\begin{equation}\label{lg1}
f= 
\frac{1}{2m }|(-i\hbar \nabla - 2e\vec{ A})
{\mathbf{\Psi}}|^{2 } + a
|{\mathbf{\Psi} }|^{2}
+ b |{\mathbf{\Psi} }^{*}\cdot{\mathbf{\Psi} }
|^{2} +d
|{\mathbf{\Psi} }\cdot {\mathbf{\Psi} }|^{2},
\end{equation}
where $\vec{A}$ is the vector potential, minimally
coupled to the order parameter. 
Provided $d>0$, 
the condensate energy is minimized when ${\mathbf{\Psi} }\cdot
{\mathbf{\Psi}} =0$ and the real and imaginary parts of the 
order parameter are orthogonal 
${\bf\Psi}  = |\Psi  | (\hat {\rm {\bf l}}    + i
\hat {\bf m} )$. The triplet odd-frequency order parameter thus defines
a triad $(\hat {\bf l } , \hat {\bf m } , \hat {\bf
n})$ of orthogonal vectors 
with principal axis $\hat{ \bf n} = \hat {\bf l } \times \hat
{\bf m }$. 

\begin{figure*}[t]
\centering
\includegraphics[width=15cm]{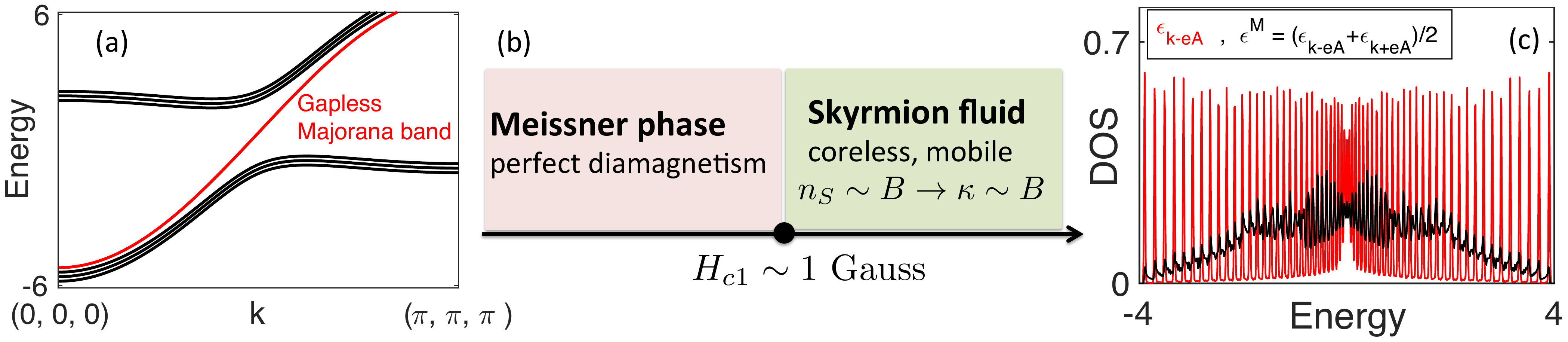}
\caption{ (a) Hybridization of 3 localized Majorana fermions per spin
with 4 Majorana fermions of the conduction band leads to one gapless
Majorana Fermi surface. (b) Magnetic field phase diagram of a Skyrme insulator.
(c) Landau quantization of the projected Majorana Fermi surface. 
}
\label{Fig2}
\end{figure*}

Eliminating the amplitude degrees of freedom
(see supplementary material)\cite{Coleman94,Knigavko_PRB1999},
the long-wavelength action has the following form
\begin{eqnarray}\label{free}
{\cal F} = \int d^{4}x\left[
\frac{\rho_{\perp}}{2}(\partial_{\mu}\hat{\bf n})^2 +
\frac{\rho_{s}}{2}(\omega_{\mu} - q A_{\mu})^2 +
\frac{F_{\mu\nu}^{2}}
{16\pi} 
 \right].
\end{eqnarray}
Here $q = 2e/\hbar $, and we have adopted the relativistic
limit of the action to succinctly include both electric and magnetic
fields\cite{footnote}, 
using the  Minkowski signature
($x_{\mu}^{2}\equiv \vec{x}^{2}- x_{0}^{2}$ with $c=1$) and
denoting $A_{\mu}= (-V,\vec{A})$ as the four-component 
vector potential. 
The first two terms describe the condensate action, 
where $\omega_{\mu}=
\hat{\bf m}\cdot \partial_{\mu}\hat{\bf l}$ is the
rate of precession of the order parameter about the 
$\hat {\bf n}$ axis. $\rho_{s}$ is the nominal superfluid stiffness, while
$\rho_{\perp }$ determines the magnetic rigidity.   
The last term is the field energy, where $
F_{\mu\nu}= \partial_{\mu}A_{\nu}-\partial_{\nu }A_{\mu} 
$ is the electromagnetic field tensor. 
The stiffness coefficients $\rho_{\perp}, \rho_{s}$ are
temperature dependent and are obtained by integrating out the thermal
and quantum fluctuations of the microscopic degrees of freedom. 
Under the gauge
transformation 
$(\bl+i \bM )\rightarrow e^{i\phi } (\bl + i \bM)$ and $qA_{\mu}
\rightarrow q A_{\mu}+ \partial_{\mu}\phi $, 
the vectors $\bl$ and $\bM $ rotate through an angle $\phi $ about the
$\bn  $ axis,
so the angular gradient
transforms as $\omega_{\mu}\rightarrow \omega_{\mu }+
\partial_{\mu}\phi $, and thus the currents ${J}^{\mu}= q
\rho_{s} (\omega^{\mu }- q {A}^{\mu})$ and 
free energy are gauge invariant. The equivalence of electron gauge
transformations and spin-rotation means that gauge transformations are
entirely contained within the $SO(3)$ manifold of the order parameter. 

To analyze how the superflow is destablized, we examine the screening
of electromagnetic fields.
From Amp\`eres equation
$4\pi J^{\mu}= \partial_{\nu}F^{\mu \nu}$,
we observe if $ \partial_{\nu}F^{\mu \nu}=0$, corresponding to 
uniform  internal fields, then the supercurrent vanishes
${J}^{\mu}= q \rho_{s} (\omega^{\mu }- q {A}^{\mu})=0$.
In a superconductor, this condition is only be achieved by the
complete exclusion of fields, 
but here the texture of the composite order parameter 
is able to continually adjust with the vector potential so that $\omega^\mu=q
A^\mu$, enabling the current to vanish in the presence of  internal 
fields.
To examine this further, we take the curl of
Amp\`eres equation,  to obtain 
\begin{equation}\label{London}
(1 -\lambda_{L}^{2}\partial^{2})F^{\mu \nu}=
q^{-1}\Omega^{\mu \nu},
\end{equation}
where $\lambda_{L} = ( 4 \pi q^{2}\rho_{s})^{-1/2}$ is the London penetration
depth. This modified London equation contains the additional term
$\Omega^{\mu \nu}= \partial^{\mu} \omega^{\nu}-\partial^{\nu}\omega^{\mu}$,
which is the curl of the gradient of the order parameter. In a conventional
superconductor, $\omega^{\mu}= \partial^{\mu}\phi $ is  the
gradient of the superconducting phase so $\Omega^{\mu\nu}=0$ vanishes
causing fields to be expelled. 
However in a Skyrme insulator, the 
quantity $\Omega^{\mu\nu}$ is finite, and can be written in
the form ${\Omega^{\mu\nu}}
 = {\bn}\cdot
(\partial^{\nu}{\bn }\times \partial^{\mu}{\bn })
$, which is a the Mermin-Ho relation\cite{SM} 
for the skyrmion density of the $\bn $ field.
From (\ref{London}), we see that on scales long compared with
the penetration depth, where
gradients of the field can be neglected,  the average skyrmion
density locks to the average external field, $
\overline{\Omega^{\mu
\nu}}=q \overline{F^{\mu\nu}}$, where the lines denote a
coarse-grained average. This relation expresses the screening
of supercurrents by the skyrmions, and it 
also holds 
in non-relativistic versions of this theory\cite{footnote}. 
Moreover, phase  rotations 
around the $\bn $ axis 
are now absorbed into the electromagnetic field (Anderson Higg's
effect), leaving behind 
a residual order parameter manifold with $SO (3)/U (1) \equiv
S^{2}$ symmetry.  
While the homotopy analysis yields no stable vortices
$\pi_1(S^2)=0$, it does allow for the topologically stable skyrmion solutions
$\pi_2(S^2)=\mathbb{Z}$ that screen the superflow and allow
penetration of electric and magnetic fields. 
We shall actually consider lines of skyrmion , 
formed by stacking two dimensional skyrmion
configurations, similar to vortex lines in three dimensional superconductors.
We call the corresponding dielectric a ``Skyrme insulator''.

Written in non-relativistic language, the equations
relating the skyrmion density to the penetrating fields are 
\begin{eqnarray}\label{thestuff}
\frac{1}{2\pi}
\overline{
 {\bn}\cdot
(\partial_{i}{\bn }\times \partial_{j}{\bn }) }
&=&-
\epsilon_{ijk}
\left(\frac{B_{k}}{\Phi_{0}} \right)\cr
\frac{1}{2\pi}
\overline{
 {\bn}\cdot
(\partial_{i} {\bn }\times \partial_{t}{\bn })} &=&
\frac{2e}{h}
{E_{i}}
\end{eqnarray}
where  $\Phi_{0} = 2\pi/q = h/2e$ is the flux quantum, and the
overline denotes a coarse-grained average over space or time.
The first term in (\ref{thestuff}) relates the areal density of
skyrmions to 
the magnetic field, allowing a magnetic field to penetrate with a
density of one flux quantum per half-skyrmion or ``meron''. 
The second term in (\ref{thestuff}) describes the unravelling of
supercurrents due to phase slippage\cite{Anderson:1977de} created by
domain wall or instanton configurations of the order parameter. The
integral of this term 
over a time $t$ and length $L$ of the wire, counts the number of
domain-walls 
$N = -\frac{2e}{h} (V_{2}-V_{1})t $ crossing the wire in time $t$, 
in the presence of a finite voltage drop $V_{2}-V_{1}$. This voltage
generation mechanism is similar to the 
development of insulating behavior in disordered two-dimensional
superconductors\cite{mpafisher}. 
We conclude that the failure of the superconductivity does not
reinstate a metal, which would screen out electric fields, 
but instead transforms it into a dielectric into which both electric
and magnetic fields freely penetrate. 

Unlike vortices, skyrmions are coreless, with short-range interactions, 
so we expect them to
form an unpinned liquid, analogous to the vortex liquid of type II
superconductors, which restores
the broken $U (1)$ symmetry on macroscopic scales. 
How then would we distinguish a Skyrme insulator from a more
conventional dielectric?
Since the density of merons (half skyrmions) $n_{s} = B/\Phi_{0}$ 
is proportional to a
magnetic field, one signature of a skyrmion liquid
is a thermal conductivity $\kappa\propto H$ 
proportional to the applied field $H$.  In a Drude
model, the drift velocity $v_{d}= \mu (-\nabla T)$ 
is proportional to the temperature gradient and the skyrmion mobility
$\mu$. If ${\cal Q}$ is the heat content per unit length, then 
$\kappa = {\cal Q}\mu n_{S} $, so that 
\begin{equation}\label{thermal}
\kappa = \left(\frac{\mu {\cal Q}}{\Phi_{0}}\right) H.
\end{equation}
is proportional to the applied field. 

A further consequence 
is the 
development of a low field Meissner phase. 
In a fixed external magnetic field ${\bf H}$, 
we consider the Gibb's free energy 
${\cal G} = {\cal F} - \int
d^{3}x {\bf H}\cdot {\bf B } (x) / (4\pi)$. 
Taking the field to lie in the z-direction and
re-writing the field $B_{z}= n_{S} (x)\Phi_{0}$, where
$n_{S}= \frac{1}{2\pi}\Omega^{12}$  is the areal meron  density, then 
\begin{eqnarray}\label{free}
{\cal G} = \int d^{3}x\left[
\frac{\rho_{\perp}}{2}(\partial_{\mu}{\bf n})^2 +
\frac{(H- \Phi_{0}{ n_{S}} (x))^{2}}{8\pi} 
-\frac{H^2}{8 \pi}
 \right],
\end{eqnarray}
This corresponds to an $O(3)$ sigma model in which the skyrmions have 
a finite chemical potential $\mu_{S} = \Phi_{0}H/4\pi$,
per unit length.  Suppose the corresponding energy of 
a skyrmion is $\epsilon_{S}/a$ per unit length, 
where $a$ is the lattice spacing, then providing $H<
H_{c}= 4 \pi \epsilon_{S}/\Phi_{0}a$, the skyrmion energy will exceed
the chemical potential, and they will be excluded from the fluid. Reverting to 
SI notation, this becomes 
\begin{equation}\label{}
\mu_{0}H_{c} = \frac{4}{137} \left(\frac{V_{S}}{a c} \right), 
\end{equation}
where we have replaced $\frac{e^{2}}{\hbar c} = 1/137$, the fine
structure constant and $\epsilon_{S}= eV_{S}$. Below this field, 
skyrmions and field lines will be expelled, so 
the material will exhibit a Meissner effect.
A generic phase diagram is given in Fig. \ref{Fig2}(b).

We now discuss the possible microscopic origin of this kind of order,
and its possible application to SmB$_{6}$. 
Various anomalous aspects of insulating SmB$_{6}$ can be speculatively
associated with the properties of a Skyrme insulator. 
The recent observation of an unusual thermal
conductivity in insulating SmB$_{6}$, 
that is linear in field, $\kappa \propto
H$\cite{Sebastian_MarchMeeting2016} 
is most naturally interpreted as a 
kind of flux liquid expected in such a phase, a hypothesis
that could be checked by confirming that this anomalous thermal
conductivity is only exhibited perpendicular to the field direction.

A second test of this hypothesis, is the magnetic susceptibility. 
In a heavy fermion compound, the order parameter stiffness $\rho $ is
set by the Kondo temperature $T_{K}$, $\rho \sim k_{B}T_{K}/a$\cite{Coleman94},
where $a$ is the lattice spacing, so the
the energy of a skyrmion is approximately
$k_{B}T_{K}$ per unit lattice spacing $a$ and $eV_{K}\sim
k_{B}T_{K}$. For SmB$_{6}$ we estimate 
$V_{K}= 1 meV$, and with $a=10^{-9}m$
we obtain $\mu_{0}H_{c}\sim 10^{-4}T$ or 1 Gauss, comparable with the
earth's magnetic field. In a magnetically 
screened ($\mu-$ metal) environment we expect 
SmB$_{6}$ to become fully diamagnetic with magnetic susceptibililty $\chi =-1/4\pi$.

A microscopic model for the development of composite order in a 
Kondo lattice was studied by Coleman, Miranda and Tsvelik\cite{Coleman94,Coleman_PhysicaB1993} (CMT)
and recently revisited by Baskaran\cite{Baskaran_arxiv2015}.
This model allows us to pursue the consequences of the 
failed-superconductivity hypothesis into the microscopic domain.
In a conventional Kondo lattice the local moments
fractionalize into charged Dirac fermions; the 
CMT model considers an alternative
fractionalization of the local moments into {\sl Majorana fermions}.  
In the corresponding mean-field theory, spin $1/2$ local moments $\bf S$ 
are represented as a bilinear of ${\bf S}=-\frac{i}{2} \boldsymbol{\hat \eta} \times \boldsymbol{\eta}$,
where $ \boldsymbol{\hat \eta} =
(\hat \eta_x, \hat \eta_y, \hat \eta_z)$ is a triplet  of Majorana fermions. 
In this representation, the Kondo interaction factorizes as  follows:
\begin{eqnarray}
H_K[i] &=& J_K (\hat \psi_{i\alpha}^\dagger \boldsymbol{\sigma}_{\alpha \beta}
\hat \psi_{i\beta})\cdot {\bf S}_i  
 \nonumber\\
&\rightarrow&\biggl[\hat  \psi_{i\alpha}^\dagger (\boldsymbol{\sigma}_{\alpha \beta} \cdot
\hat {\boldsymbol \eta}_i) {\cal V}_{i\beta} +{\rm H.c}\biggr] + { {\cal V}_{i}\dg {\cal V}_{i}}/J_{K},
\end{eqnarray}
where $J_K$ is the Kondo interaction strength, 
$c\dg _{i \gamma}$ creates a conduction electron 
and 
$[{\cal V}_{i}]_{\beta}
=-\frac{J_K}{2}
\langle(\boldsymbol{\sigma}_{\beta
 \gamma} \cdot \boldsymbol{\eta}_i ) c_{i \gamma}\rangle $ is a
 two-component spinor. ${\cal V}_{j}$ 
determines the composite order via the  equation $\vec{\Psi } ({\bf x}) = {\cal V}^{T} i\sigma_{2}\vec{\sigma }{\cal V}$.
We have 
extended the CMT model to include spin-orbit coupling by incorporating 
a `p-wave' form factor into the definition of the conduction Wannier
states $c_{i}$, derived from the angular momentum difference $|\Delta
l | = 1$ between the heavy $f$ and light $d$ electrons\cite{SM, Alexandrov_PRL2015}.
Our mean-field calculations confirm that even in the presence of the spin-orbit
coupling, the ground-state energy 
is independent of the 
 orientation of the composite order parameter $\vec{\Psi }$, 
so the system remains isotropic\cite{SM}.

In the CMT model, the conduction electrons, represented by four degenerate
Majorana bands, hybridize with the three neutral Majorana
fermions, gapping all but one of them which is left behind 
to form a gapless Majorana Fermi sea [Fig \ref{Fig2} (a)]. 
This unique feature provides an
appealing explanation of the robust linear specific heat $C_{v}=\gamma
T$ observed in this material.
The neutrality of the Majorana Fermi sea eliminates the strictly DC
conductivity,  but the current and spin matrix elements 
are actually proportional to energy,
which will lead to a quasiparticle optical conductivity 
of the form 
\begin{align}
{\rm Re}  [\sigma(\omega)]=\frac{\sigma_0}{1+\omega^2 \tau^2} \omega^2,
\end{align}
where $\tau$ is the relaxation rate.
The analogous matrix element effect also suppresses the Koringa spin
relaxation rate, giving rise to 
a $T^3$ NMR relaxation rate\cite{Coleman_PhysicaB1993}.   When we include the spin-orbit
coupling, we find that an additional topological Majorana surface
state develops, reminiscent of the Majorana surface states of
superluid He-3. This interesting state is protected by the crystal mirror
symmetry and decouples from the  gapless bulk band. 
Thus the insulating state retains some of the surface
conductivity of a topological Kondo insulator\cite{Dzero_PRL2010,Wolgast_PRB2013}. 

Perhaps the most puzzling aspect of SmB$_{6}$ is the reported observation of 
3D bulk quantum oscillations.  An approximate treatment of the effect
of a magnetic field on the Majorana Fermi surface 
can be made by initially ignoring the  
skyrmion fluid background. 
The dispersion of the Majorana band in a field
can then be calculated by projecting the
Hamiltonian into the low-lying Majorana band.
\begin{align}
\epsilon^{\rm M}_{\bf k, A}=\langle \phi^{\rm M}_{\bf k}|H({\bf k,A})|\phi^{\rm M}_{\bf k}\rangle =\frac{1}{2}(\epsilon^{\rm e}_{{\bf k}-e{\bf A}}+\epsilon^{\rm h}_{{\bf k}+e{\bf A}}),
\end{align}
where $\epsilon^{\rm e}_{\bf k-eA}$ and $\epsilon^{\rm h}_{\bf k+eA}$
are the dispersion for electrons and holes which couple to the
external gauge field with opposite signs.   Although the scattering off the
triplet condensate mixes the electron and hole components of the
field, giving rise to a neutral quasiparticles for which
current operator $J_\alpha = \partial \epsilon^{\rm M}_{\bf k,A} /\partial
A_\alpha|_{\bf A=0}=0$ vanishes, this cancellation does not
extend to the second 
derivative of the energy $\partial^2 \epsilon^{\rm M}_{\bf k,A} /\partial { A_\alpha}^2|_{\bf A=0}\neq
0$ which is responsible for the diamagnetic response. 
This is a consequence of the broken gauge-invariant
environment provided by the Skyrme insulator. 
In
Fig. \ref{Fig2}(c), we show the density of states of the Majorana band
in a magnetic field, demonstrating a discrete Landau quantization
with broadened Landau levels.  Since quantum oscillations
originate from the discretization of the density of states
into Landau levels, we anticipate that a Majorana Fermi surface does give rise to quantum
oscillations. Moreover since the Majorana Fermi surface originates
predominantly from the conduction electron band, it has a small
effective mass, in accordance with quantum oscillation
experiments\cite{Li_Science2014, Sebastian_Science2015}.


We note that triplet odd frequency pairing 
is expected to be highly prone to disorder. 
Weakly
disordered samples may indeed revert to a topological Kondo insulating  phase, 
at least in the majority of the sample.  This may account for the
marked sample dependence, and the discrepancies between samples grown
by different crystal growth techniques. 
Nevertheless, we expect that
small patches of failed superconductivity will still lead to 
enhanced diamagnetism in a screened ($\mu-$ metal) environment.

Our results also set the stage for a broader consideration 
of failed superconductivity in other strongly correlated materials. 
There are several known Kondo insulators 
with marked linear specific heat coefficients, including 
Ce$_3$Bi$_4$Pt$_3$\cite{Jaime_Nature2000}, 
CeRu$_4$Sn$_6$\cite{Bruning_PRB2010} and
CeOs$_4$As$_{12}$\cite{Baumbach_PNAS2008} which might fall into this class. 
We end by noting that Skyrme insulators may also be relevant in
an astrophysical context such as color superconductivity in
white dwarf or neutron stars\cite{Ginzburg,RevModPhys.80.1455}.

We thank Peter Armitage, 
Eric Bauer, Michael Gershenson, Andrew Mackenzie, Filip Ronning and Suchitra Sebastian
for useful discussions related to this work.  This work was
supported by the Rutgers Center for Materials Theory group postdoc
grant (Po-Yao Chang), Piers Coleman and Onur Erten were supported
by the U.S. Department of Energy basic energy sciences grant
DE-FG02-99ER45790. Piers Coleman and Onur Erten also acknowledge the hospitality of 
the Aspen Center for Physics, which is supported by National Science Foundation grant PHY-1066293. 
Alexei Tsvelik was supported by the
U.S. Department of Energy (DOE), Division of Condensed Matter
Physics and Materials Science, under Contract No. DE-AC02-98CH10886.

%

\clearpage

\pagebreak

\newpage

\begin{center}

\noindent{\Large Supplementary Material for ``Skyrme insulators:
insulators at the brink of superconductivity. 
" }
\end{center}
\setcounter{equation}{0}
\setcounter{figure}{0}
\setcounter{table}{0}
\setcounter{page}{1}
\setcounter{section}{0}


\bigskip
These supplementary materials describe the details behind the
phenomenology of a Skyrme insulator, and the underlying
microscopic mean-field theory.

\tableofcontents
\section{Phenomenology}

\subsection{Ginzburg-Landau theory}

To illustrate the idea of a Skyrme insulator, we consider a
superconductor with a complex vector order parameter $\vec{\Psi
}$.  In our microscopic realization of this phenomenon, the complex vector
order parameter is a consequence of underlying odd-frequency triplet
pairing, which gives rise to composite order 
between the pair density of a conduction sea, and the staggered
magnetization of a Kondo lattice, given by 
\begin{equation}\label{}
\vec{\Psi } ({\bf x})= 
\left(-1 \right)^{i+j+k}
\langle \psi_{\uparrow}\psi_{\downarrow } {\bf S} ({\bf x})\rangle  =
{\frac{\vert \vec{\Psi }({\bf x})\vert}{\sqrt{2}}} \biggl(
\hat {\rm {\bf l}} ({\bf x})   + i   \hat {\bf m} ({\bf x})\biggr),
\end{equation}
where 
$\psi_{\uparrow}\psi_{\downarrow }$ is the pair density and 
$
\left(-1 \right)^{i+j+k}{\bf S} ({\bf x})$ is the staggered
magnetization.  

However, the long-wavelength action can be developed
independently of the microscopic theory 
by considering the Landau Ginzburg theory of a
complex vector $\vec{ \Psi }$.  The first part of our derivation closely 
follows reference\cite{Knigavko_PRB1999}. 
Considering terms up to quartic order,
the most general isotropic Landau Ginzburg free energy functional of a 
complex three component vector $\vec{\Psi }$ is given by
\begin{equation}\label{}
{\cal F}[\Psi] = \int d^{3}x f[\Psi],
\end{equation}
where 
\begin{eqnarray}\label{l}
f[\Psi ]&=& \frac{\hbar^{2}}{2m }|(\partial_{j} + i q {A}_{j})
\vec{\Psi}|^{2 } + a
|\vec{\psi }|^{2}\cr
&+& b |\vec{\Psi }^{*}\cdot\vec{\Psi }|^{2} + d 
|\vec{\Psi }\cdot \vec{\Psi }|^{2}
+e|\vec{\Psi }\times \vec{\Psi}^{*}|^{2},
\end{eqnarray}
where $q = 2 e / (\hbar c)$ and summation over $j\in [1,3]$ is implied.
By re-writing 
$|\vec{\Psi }\times \vec{\Psi}^{*}|^{2}
=
|\vec{\Psi }|^{4}
- |\vec{\Psi }\cdot \vec{\Psi }|^{2}$, and changing 
$b\rightarrow b-e$, $d\rightarrow d+e$, we can absorb the last quartic
term into redefinitions
of $b$ and $d$, so that 
\begin{equation}\label{lg1}
f[\Psi]= 
\frac{\hbar^{2}}{2m }|(\partial_{j} + i q {A}_{j})
\vec{\Psi}|^{2 } + a
|\vec{\Psi }|^{2}
+ b |\vec{\Psi }^{*}\cdot\vec{\Psi }|^{2} + d 
|\vec{\Psi }\cdot \vec{\Psi }|^{2}
\end{equation}

Splitting the order into its real and imaginary components, 
$\vec{\Psi } = \vec{\Psi}_{1} + i {\vec \Psi}_{2}$, then
\begin{equation}\label{lg1}
|\vec{\Psi }\cdot\vec{\Psi } |^{2}
= ( |\Psi_{1}|^{2} - |\Psi_{2}|^{2})^{2}+ 4|
\vec{ \Psi}_{1}\cdot \vec{\Psi}_{2}|^{2}.
\end{equation}
If $d>0$, it follows that the energy is  minimized by configurations in
which  $\vec{\Psi }_{1}$ and $\vec{\Psi }_{2}$ are orthogonal, and of 
equal magnitude $|\Psi |$, so that $\vec{\Psi }\cdot\vec{\Psi } =0$.  
It follows that the order parameter has the general form 
\begin{equation}\label{}
\vec{ \Psi  } (x) = { \frac{|\Psi (x) |}{\sqrt{2}}} (\bl (x) +  i \bf{\hat  m} (x)).
\end{equation}
(Note that the equivalent form $ {\frac{|\Psi (x) |}{\sqrt{2}}} (\bl (x) -  i \bf{\hat
m} (x))$ can be transformed into the above by a 180$^{0}$ rotation in
spin space about the $\bl$ axis.)
In these new variables the amplitude variables separate out from the 
orientational degrees of freedom, and the Landau
Ginzburg free energy takes the following form 
\begin{eqnarray}\label{l}
f[\Psi]&=& \frac{\hbar^{2}}{2m }|(\partial_{j}|\Psi |)^{2}+ a
|\Psi |^{2}+ b |\Psi |^{4}
\cr
&+&\frac{\hbar^{2}}{2m } {\frac{ |\Psi |^{2}}{2}}
(\partial_{j} + i q {A}_{j})
(\bl  +  i \bf{\hat  m} )|^{2 } 
\end{eqnarray}
Now if $\hat {\bf n} = \bl \times \bf{\hat m}$, then $(\bl,\bM,\bn
)$ forms a right-handed basis. We can
re-write the derivatives of the basis vectors in terms of an angular
velocity $\bw $, such that $\partial_{j} (\bl,\bM,\bn) = \bw_{j}\times 
(\bl,\bM,\bn)$. It follows that:
\begin{eqnarray}\label{l}
\partial_{j} (\bl+i\bM) &=& {\bw_{j}}\times (\bl+i\bM),\cr
(\partial_{j}+i q A_{j}) (\bl+i\bM) &=& ( {\bw_{j}- q A_{j}\bn })\times (\bl+i\bM).
\end{eqnarray}
Decomposing $\bw_j = \omega_{j}^{1}\bl+\omega_{j}^{2}\bM +
\omega_{j}^{3}\bn $, we can then write
\begin{eqnarray}\label{l}
\partial_{j} (\bl+i\bM) &=&  (\omega^{2}_{j}\bM  +
\omega^{3}_{j}\bn)\times \bl + i (\omega^{2}_{j}\bl+\omega^3_{j}\bn
)\times \bM \cr
&=&
-i \omega^{3}_{j} (\bl + i \bM ) + i
(\omega^{1}_{j}+ i \omega^{2}_{j})\bn 
\end{eqnarray}
and hence
\begin{eqnarray}\label{l}
(\partial_{j} - iq A_{j} )(\bl+i\bM) &=&
-i (\omega^{3}_{j} -q A_{j})(\bl + i \bM ) \cr
&+& i
(\omega^{1}_{j}+ i \omega^{2}_{j})\bn 
\end{eqnarray}
and hence
\begin{eqnarray}\label{l}
|(\partial_{j} + i q {A}_{j})
(\bl  +  i \bM )|^{2 }  &=& \biggl ((\omega^{1}_{j})^{2}+
(\omega^{2}_{j} )^{2}\biggr) + 2 (\omega^{3}_{j}-q A_{j})^{2}\cr
&=& (\partial_{j}\bn )^{2} + 2 (\omega^{3}_{j}-q A_{j})^{2}.
\end{eqnarray}
Using these results, the Landau Ginzburg free energy can now be written
as
\begin{eqnarray}\label{l}
f[\Psi]&=& \frac{\hbar^{2}}{2m }|(\partial_{j}|\Psi |)^{2}+ a
|\Psi |^{2}+ b |\Psi |^{4}
\cr
&+& \frac{\rho_{\perp }}{2}(\partial_{j}\bn )^{2} + \frac{\rho_{s}}{2} (\omega^{3}_{j}-q A_{j})^{2}.
\end{eqnarray}
where the stiffnesses are given by 
\begin{equation}\label{}
\rho_{s}= 2\rho_{\perp }  = \frac{\hbar^2|\Psi|^2}{m}
\end{equation}
If we now neglect the amplitude terms, the final Landau free energy
takes the form 
\begin{eqnarray}\label{l}
{\cal F}=\int d^{3}\left[
\frac{\rho_{\perp }}{2}(\partial_{j}\bn )^{2} + \frac{\rho_{s}}{2} (\omega^{3}_{j}-q A_{j})^{2} \right].
\end{eqnarray}

\subsection{Free energy and action}

To go from the free energy functional to the to the action, we now
add time-dependent quadratic terms
into the free energy functional, writing
\begin{eqnarray}
\label{free-non}
{\cal F} =& \int d^{3}xdt
\frac{1
}{2}\biggl[\rho_{\perp}(\partial_{i}\hat{\bf n})^2 -
\chi_{\perp}(\partial_{t}\bn )^{2}
 \biggr]   \notag \\
&+ \frac{1}{2}\biggl[\rho_{s} (\omega^3_{i} - q A_{i})^2 
- \chi_{s}
(\omega^3_{0} - q A_{0})^2 
\biggr]
\end{eqnarray}
Here $A_{\mu}= (A_{0},\vec{A}) = (- c V , \vec{A}) $ is the four-component
vector potential, where $c$ is the speed of light and $V$ is the
scalar potential. 
The quantities $\chi_{\perp }$  and $\chi_{s}$ are the
susceptibilities of the order parameter. In the absence of an
electromagnetic field, these give rise to 
Bolguilubov and spin-waves mode with respective velocities
$c_{s}= (\rho_{s}/\chi_{s})^{\frac{1}{2}}$ and 
$c_{\perp }= (\rho_{\perp }/\chi_{\perp })^{\frac{1}{2}}$.
If we now include the action of the electromagnetic field, we obtain
\begin{eqnarray}\label{free-non}
{\cal F} &=& \int d^{3}xdt
\left ( 
\frac{\rho_{\perp}}{2}\biggl[(\partial_{i}\hat{\bf n})^2 -
\frac{1}{c_{\perp }^{2}} (\partial_{t}\bn )^{2}
 \biggr]\right.\cr
&+&
\left. \frac{\rho_{s}}{2}\biggl[ (\omega^3_{i} - q A_{i})^2 
- \frac{1}{c_{s}^{2}}
(\omega^3_{0} - q A_{0})^2 
\biggr]
+ \frac{{\bf B}^2- {\bf E}^2}
{8\pi}\right )
,\cr
&&
\end{eqnarray}
where we use cgs units. 
This is the non-relativistic version of the theory. 

For ease of analysis, it is useful to consider the relativistic limit
of this action.  Departures from relativistic behavior can easily be
added back in at a later stage.  In the relativistic version of the
theory, $c_s=c_{\perp} =c $, the speed of light. 
In this case, we can rescale the time-components, writing 
$x^{0}= c t $, so that 
$\frac{1}{c}\partial_t \to \partial_0$
and $\frac{1}{c}(\omega_0-q A_0)\to (\omega_0-q A_0)$, where now 
$A_{\mu}\equiv (A_{0},\vec{A})= (-V,\vec A)$

The long-wave length action then acquires the manifestly relativistic form
\begin{eqnarray}\label{free-rel}
{\cal F} =&
\int d^{4}x
\left[
\frac{\rho_{\perp}}{2}
(\partial_{\mu}\hat{\bf n})(\partial^{\mu}\hat{\bf n})+
\frac{\rho_{s}}{2}(\omega^3_{\mu} - q A_{\mu})(\omega^{3\mu} - q A^{\mu}) \right]  \notag\\
&+\frac{(F_{\mu\nu})^2}
{16\pi},
\end{eqnarray}
where we have adopted a relativistic notation with  Minkowski metric
$g_{\mu\nu}=(-1,1,1,1)$ and $F_{\mu\nu} = \partial_{\mu}A_{\nu}
-\partial_{\nu}A_{\mu} $ is the electromagnetic  tensor, such that
$F_{i0}=E_{i}$ and $F_{ij}=\epsilon_{ijk}B_{k}$  determine the
electric and magnetic fields respectively. 

To determine Maxwell's equations in the presence of the order
parameter, we take variations with respect to $\delta A^{\mu}$, 
\begin{equation}\label{}
\delta {\cal F} = \int d^{3 }xdt\left[- J_{\mu} + \frac{1}{4\pi}\partial^{\nu}F_{\mu\nu} \right]\delta A^{\mu}
\end{equation}
where $J_{\mu}= q\rho_{s} (\omega_{\mu}^{3}-q A_{\mu})$ is the
current, so that the Amp\'eres equation is 
$\partial^{\nu }F_{\mu\nu}
= 4 \pi J_{\mu} = 4 \pi q\rho_{s} (\omega_{\mu}^{3}-q A_{\mu})$. In a
conventional superconductor $\omega_{\mu}^{3} = \partial_{\mu}\phi $
is just the gradient of a phase, a quantity that can not develop a
rotation, and (without vortices) 
this prohibits solutions in which the magnetic field
uniformly penetrates the sample. 
However, for this SO(3) order parameter,
defined by three Euler angles $(\phi , \theta ,\psi )$,
$\omega_{\mu}^{3} = \partial_{\mu}{ \psi} +  {\cos
\theta} \partial_{\mu }{\phi} $ can develop a non-zero curl. Taking the curl of Amp\'eres
equation we obtain 
\begin{eqnarray}\label{l}
(\partial_{\mu}J_{\nu}- \partial_{\nu }J_{\mu }) &=& q\rho_s
(\omega_{\mu\nu}-q F_{\mu\nu}) \cr 
&=& \frac{1}{4\pi}
\left[\partial_{\mu}\partial^{\eta} F_{\nu\eta }- (\mu \leftrightarrow
\nu)
\right]\cr
&=& \frac{1}{4\pi}
\left[
\partial_{\mu}\partial^{\eta} 
(\partial_{\nu} A_{\eta } - \partial_{\eta }A_{\nu})
- (\mu \leftrightarrow
\nu)
\right]\cr
&=& - \frac{1}{4\pi} \partial^{2}F_{\mu\nu}
\end{eqnarray}
where 
$\omega_{\mu\nu} = 
\partial_{\mu}\omega^{3}_{\nu }-
\partial_{\nu}\omega^{3}_{\mu }$ 
is the curl of the rotation. Rearranging this result gives the London
equation 
\begin{equation}\label{}
\lambda_{L}^{2}\partial^{2}F_{\mu\nu} = F_{\mu\nu} - q^{-1}\omega_{\mu\nu}
\end{equation}
where the penetration depth $\lambda_{L} = (4 \pi
q^{2}\rho_{s})^{-1/2}$. To seek uniform solutions, we set the
left-hand side to zero.  Using the Mermin-Ho relation, (see Sec. \ref{Sec:MH})
$\omega_{\mu\nu} =  - \bn \cdot(\partial_{\mu}\bn
\times 
\partial_{\nu}\bn), 
$
when a uniform field penetrates the material, we obtain 
\begin{equation}\label{}
\frac{\bn  \cdot(\partial_{\mu}\bn \times \partial_{\nu}\bn )}{2\pi}=-\frac{F_{\mu\nu} }{\Phi_{0}}
\end{equation}
where  $\Phi_{0} = \frac{2\pi}{q}= \frac{h}{2e}$ is the
superconducting flux quantum, in SI units. Identifying $F_{ij}=
\epsilon_{ijk}B_{k}$ and $F_{i0}= E_{i}$ this equation can be written
as two separate equatins relating the penetrating fields to the
skyrmion density,
\begin{eqnarray}\label{thestuff}
\frac{1}{2\pi}
 {\bn}\cdot
(\partial_{i}{\bn }\times \partial_{j}{\bn }) &=&-
\epsilon_{ijk}
\left(\frac{B_{k}}{\Phi_{0}} \right)\cr
\frac{1}{2\pi}
 {\bn}\cdot
(\partial_{i}{\bn }\times \partial_{t}{\bn }) &=&
\frac{2e}{h}{E_{i}}.
\end{eqnarray}

\subsection{Derivation for non-relativistic case}\label{}

We now repeat the above derivation for the non-relativistic case. If
we restore the distinction between $c_{s}$, $c_{\perp }$ and the speed
of light, the non-relativistic action is written
\begin{eqnarray}\label{free-non}
{\cal F} &=& \int d^{4}x
\left ( 
\frac{\rho_{\perp}}{2}\biggl[(\partial_{i}\hat{\bf n})^2 -
\frac{c^{2}}{c_{\perp }^{2}} (\partial_{0}\bn )^{2}
 \biggr]\right.\cr
&+&
\left. \frac{\rho_{s}}{2}\biggl[ (\omega^3_{i} - q A_{i})^2 
- \frac{c^{2}}{c_{s}^{2}}
(\omega^3_{0} - q A_{0})^2 
\biggr] 
+ \frac{F_{\mu \nu}^{2}}
{16\pi}\right )
,\cr
&&
\end{eqnarray}
where $x_{0}\equiv  c t$ as in the relativistic case. We see that the
modification to the velocities only affects the temporal components of
the action.  When we take the variation with respect to $A^{\mu}$ the
relativistic Amp\'ere equation now contains a correction to the zeroth
order term, 
\begin{equation}\label{}
4 \pi q\rho_{s} (\omega_{\mu}^{3}-q A_{\mu})= \partial^{\nu }F_{\mu\nu}
+ \left(\frac{c_{s}^{2}}{c^{2}}-1 \right) \delta_{\mu 0} \partial^{\nu }F_{0\nu}
\end{equation}
Taking the curl of this equation and identifying $F_{0i}=-E_{i}$we obtain
\begin{align}
\label{}
&(F_{\mu\nu}- q^{-1}\omega_{\mu\nu})  \notag \\
=&\lambda_{L}^{2}
\left(\partial^{2}F_{\mu\nu}+\left[\frac{c_{s}^{2}}{c^{2}}-1
\right] (\delta_{\nu 0}\partial_{\mu}- \delta_{\mu 0} \partial_{\nu})
\vec{\nabla} \cdot \vec{ E}
 \right).
\end{align}
The magnetic part of this equation, $F_{ij}- q^{-1}\omega_{ij}=
\lambda_{L}^{2}\partial^{2}F_{ij}$, ($i,j\in [1,3]$) is unaltered.  
Identifying $F_{i0}=E_{i}$, the electric part now reads
\begin{equation}\label{}
E_{i}-q^{-1}\omega_{i0}= \lambda_{L}^2\left[\partial^{2}E_{i} + \left[\frac{c_{s}^{2}}{c^{2}}-1
\right]\partial_{i}\vec{\nabla} \cdot \vec{ E} \right].
\end{equation}
If we decouple the electric field into its longitudinal and transverse
components $\vec{E}= \vec{E}^L+ \vec{E}^{T}$
{with $\vec{\nabla} \cdot \vec{E}^T=0$ and $\vec{\nabla} \times \vec{E}^L=0$ }, this becomes
\begin{align}\label{}
E_{i}-q^{-1}\omega_{i0}  =\lambda_{L}^2\left[ \partial^{2}\vec{E}^{T}_i +
\frac{c^{2}_{s}}{c^{2}}\left({\partial_i (\vec{\nabla} \cdot\vec{E}^{L})} -
\frac{1}{c_{s}^{2}}\frac{\partial \vec{E}^{L}_i}{\partial t^{2}} \right)
\right].
\end{align}
The main effect of the non-relativistic equation is to renormalize the
velocity $c\rightarrow c_{s}$ and penetration depth
$\lambda_{L}\rightarrow \lambda_{T}= (c_{s}/c)\lambda_{L}$ 
of longitudinal fields. 
Once again however, the
only uniform solutions must satisfy $F_{\mu\nu}= q^{-1}\omega_{\mu \nu}$.

\subsection{Mermin-Ho relation}\label{Sec:MH}
For completeness, here we give the derivation of the Mermin-Ho
relation
\begin{equation}\label{mho}
\omega_{\mu\nu} = 
\partial_{\mu}\omega^{3}_{\nu }-
\partial_{\nu}\omega^{3}_{\mu } = - \bn \cdot(\partial_{\mu}\bn
\times 
\partial_{\nu}\bn).
\end{equation}
To derive this standard result, we write 
$\omega^{3}_{\mu} = \bM \cdot \partial_{\mu} \bl$, so that 
\begin{eqnarray}\label{l}
\partial_{\mu}\omega_{\nu}^{3}- 
\partial_{\nu}\omega_{\mu}^{3}= 
\partial_{\mu} \bM  \cdot\partial_{\nu}\bl
- 
\partial_{\nu} \bM  \cdot\partial_{\mu}\bl.
\end{eqnarray}
Expanding $\partial_{\mu} (\bl,\bM) = \bw_{\mu}\times  (\bl,\bM)
$, we obtain 
\begin{eqnarray}\label{r1}
\partial_{\mu}\omega_{\nu}^{3}- 
\partial_{\nu}\omega_{\mu}^{3}&=&
(\bw_{\mu}\times \bM )\cdot 
(\bw_{\nu}\times \bl ) - (\mu\leftrightarrow  \nu),
\cr &=& 
(\omega_{\mu}^{1}\bn  - \omega_{\mu}^{3}\bl)\cdot
(-\omega_{\nu}^{2}\bn  + \omega_{\nu}^{3}\bM)- (\mu \leftrightarrow \nu),\cr
&=& - (\omega^{1}_{\mu}\omega^{2}_{\nu}- \omega^{1}_{\nu} \omega^{2}_{\mu}).
\end{eqnarray}
By contrast, 
\begin{eqnarray}\label{r2}
\bn \cdot (\partial_{\mu}\bn  \times \partial_{\nu}\bn )&=& 
\bn \cdot\left[(\bw_{\mu}\times \bn )\times (\bw_{\nu}\times  \bn )
 \right],\cr
&=& \bn  \cdot \left[(- \omega_{\mu}^{1}\bM+\omega_{\mu}^2 \bl)\times 
(- \omega_{\nu}^{1}\bM+\omega_{\nu}^2 \bl) \right],\cr
&=& \bn \cdot \left[\omega_{\mu}^{1}\omega_{\nu}^{2}\bn 
- \omega^{2}_{\mu}\omega_{\nu}^{1}\bn  \right],\cr
&=&(\omega_{\mu}^{1}\omega_{\nu }^{2}- 
\omega_{\nu}^{1}\omega_{\mu }^{2} ).
\end{eqnarray}
Comparing (\ref{r1}) and (\ref{r2}) gives the Mermin-Ho relation
(\ref{mho}).

We can recognize the quantity  $\bn \cdot (\delta_{\mu}\bn\times
\delta_{\nu}\bn )$ as the solid angle subtended by the vector $\bn $
over the rectangle of dimensions $\delta x_{\mu}\times  \delta
x_{\nu}$. The quantity 
\begin{equation}\label{}
\frac{\omega_{\mu\nu}
}{4\pi} = \frac{\bn  \cdot(\partial_{\nu}\bn \times \partial_{\mu}\bn )}{4\pi}
\end{equation}
thus measures the areal density of skyrmions, where each skyrmion in
the order parameter encloses a solid angle $4\pi$.
On a taurus 
\begin{equation}\label{}
\int dx dy \frac{\bn  \cdot(\partial_{x}\bn \times \partial_{y}\bn
)}{4\pi} = N
\end{equation}
measures the integer number of skyrmions. 
The quantity 
\begin{equation}\label{}
\frac{\omega_{\mu\nu}
}{2\pi} = \frac{\bn  \cdot(\partial_{\nu}\bn \times \partial_{\mu}\bn )}{2\pi}
\end{equation}
measures the density of half-skyrmions, or ``merons''.

\subsection{Effective action in a magnetic field}\label{}

The purpose of this section is to immerse the Skyrme insulator in
an external magnetic field $H$ and to derive the effective long-wavelength
action that appears once the internal vector potential has been
integrated out of the dynamics.
For this calculation, we remove the time-dependent terms in the
action, reverting to the free energy. We shall also emply a simplified
two dimensional model, introducing the Free energy per layer given by  
\bea
G = a_{0}\int d^{2}x
\frac{\rho_{\perp}}{2}(\p_{\mu}{\bf n})^2 +
\frac{\rho_s}{2}(\omega^3_{\mu} - q A_{\mu})^2  \notag\\
+\frac{1}{8\pi}[\vec
\nabla\times {\bf A}]^2
- \frac{1}{4\pi} {\bf H}\cdot[\vec\nabla\times {\bf A}],
\label{free}
\eea
where $a_{0}$ is the interlayer distance, 
${\bf H}= H \hat {\bf z}$ is a uniform external field in the $z$
direction, such that ${\bf B} (x) = {\bf H}+ 4\pi {\bf M} (x)$ is the
microscopic field.  The main heuristic result, is that at
long distances, the supercurrent term $\omega^{3}_{\mu}-q
A_{\mu}\rightarrow 0$, and we can use the relationship (\ref{thestuff})
\begin{equation}\label{}
{\bf B}= \vec \nabla\times {\bf A}\rightarrow \Phi_{0}n_{s} (x)
\end{equation}
where  $\Phi_{0} = 2\pi /q= h/2e$  is the flux quantum,  and 
\begin{equation}\label{}
n_{s} (x) = \frac{1}{2\pi } {\bn  }\cdot (\partial_{y}\bn \times \partial_{x}\bn )
\end{equation}
is the meron density (each skyrmion contains two merons), to replace
the vector potential terms, so that
\begin{eqnarray}\label{free}
G = a_{0}\int d^{2}x\left[
\frac{\rho_{\perp}}{2}(\partial_{\mu}{\bf n})^2 +
\frac{(H- \Phi_{0}{ n_{S}} (x))^{2}}{8\pi} 
-\frac{H^2}{8 \pi}
 \right]. \notag\\
\end{eqnarray}
We shall actually a more general expression, valid at intermediate
distances, 
\begin{eqnarray}\label{free}
G/a_{0} &=& \int d^{2}x\left[
\frac{\rho_{\perp}}{2}(\partial_{\mu}{\bf n})^2
-\frac{H^2}{8 \pi}
 \right]\cr  &+& \frac{1}{8\pi}\int d^{2}xd^{2}x'  
(H- \Phi_{0}{ n_{S}} (\bx))
V (|\bx-\bx'|)  \notag\\
&&\times(H- \Phi_{0}{ n_{S}} (\bx')),
\end{eqnarray}
where the Fourier transform of the interaction is given by 
\begin{equation}\label{}
V (\bp) = \frac{1}{1+ {\bf p}^{2}\lambda_{L}^{2}
}
\end{equation}
where $\lambda_{L}^{2}= \frac{1}{4 \pi\rho_{s}q^{2}}$
and ${\bf p}^2=\sum_{i=1,2} p_i^2$.
When the density of 
vortices is smaller than $4\pi\rho_s$ the interaction can be
considered as short range and the model becomes an O(3) sigma model
with an additional term which
induces a skyrmion chemical potential $-H/4\pi$ and a
short-range repulsion between skyrmions.  The long-wavelength action
admits finite energy solutions in which the field penetrates and is
compensated by a finite topological charge density.

In the calculation that follows, we set $a_{0}=1$ and $q=1$. Our first
step is to gauge away the longitudinal (curl-free) parts of the
angular velocity. 
The rate of rotation around the principle axis of the order parameter
$\omega_{\mu}^{3}$ can be decomposed in terms of Euler angles:
\begin{equation}\label{}
\omega_{\mu}^3 = \p_{\mu}\psi + \cos\theta\p_{\mu}\phi. 
\end{equation}
In two dimensions, it can also be separated into a gradient and curl: 
\bea
\omega_{\mu}^3 = 
\p_{\mu}\Psi
+ \epsilon_{\mu\nu}\p_{\nu}\eta.
\eea
Since the second term is divergence free, 
this can also be thought of as a decomposition into a longitudinal and
transverse component. 
The curl of the rotation rate
\bea
\epsilon_{\mu\nu}\p_{\mu}\omega^3_{\nu} =
\epsilon_{\mu\nu}\p_{\mu}\epsilon_{\nu \gamma}\p_{\gamma}\eta
= -\nabla^2\eta = 2 \pi n_{S}\equiv \tilde{n}_{s} 
\eea
where we have defined $\tilde{n}_{S}\equiv  2 \pi n_{S}\equiv
\Phi_{0}n_{S}$.
Thus the curl of the rotation rate
is related to the density of topological charge of the ${\bf n}$ field.

Similarly, we split the vector potential into a longitudinal and transverse
 component:
 \bea
 A_{\mu} = \p_{\mu}{ \Psi} + \epsilon_{\mu\nu}\p_{\nu}\chi,
 \eea
where we have choosen a gauge to cancel the longitudinal gauge part of
$\omega_{\mu}^{3}$. 
Then the $A$-dependent term in the free energy becomes
 \bea
 (\omega_{\mu}^3 - A_{\mu})^2 = 
[(
\epsilon_{\mu \nu}\p_{\nu}(\chi-\eta)]^2 = [\partial_{\mu}(\chi -\nu)]^{2}
 \eea
while 
\begin{equation}\label{}
\vec\nabla \times \vec{A}= -\nabla^{2} \chi.
\end{equation}
The
Free energy for $H=0$ is
\begin{eqnarray}\label{absence}
G = \int d^{2} x 
\left( 
\frac{\rho_{\perp }}{2} (\partial_{\mu}\vec{n})^{2} + \frac{\rho_s}{2}[\nabla (\chi -\eta )]^{2} +\frac{1}{8\pi}  (\nabla^{2} \chi )^{2}    \right).\notag\\
\label{Eq:f}
\end{eqnarray}
 
We can rewrite the coupling between the transverse components of the
vector potential and order parameter fields {[the last two terms in Eq. (\ref{Eq:f})]}, represented by 
$\chi$ and $\eta $, in 
momentum space as follows
\begin{equation}\label{gaussian}
G_{\eta \chi }= 
\frac{1}{2}\int { \frac{{d^2p}}{(2\pi)^{2}}}\left[ 
\rho_s {\bf p}^{2} | \chi_{p}-\eta_{p}|^{2} +\frac{1}{4\pi}|{\bf p}^{2}\chi_p|^{2}
\right],
\end{equation}
where $\chi_p$ and $\eta_p$ are the Fourier components of $\chi$ and $\eta $ in the momentum space,
${\bf p}^2=\sum_{i=1,2} p_i^2$,
and $ \mathcal{G}_{\eta\chi}$ is the free energy density of the order parameter fields.
 Minimizing the free energy with respect to $\chi_{p}$ 
we then get 
\begin{equation}\label{}
\frac{\delta {\cal G}_{\eta\chi}}{\delta \chi_{p}}= 
\rho_s{\bf p}^{2} (\chi_{p}-\eta_{p})+ \frac{1}{ 4\pi}({\bf p}^{2})^2\chi_{p}=0
\end{equation}
so that 
 \bea
 \chi_{p} &=& \frac{\rho_s}{\rho_s + {\bf p}^2/4\pi}\eta_{p},
\cr
\eta_{p} -\chi_{p} &=& 
\frac{{\bf p}^{2}}{4\pi\rho_s + {\bf p}^2}\eta_{p}
=
-\frac{1}{4\pi\rho_s + {\bf p}^2}\tilde{n}_{S} (p)
 \label{sol},
 \eea
and the magnetic field is
 \bea
 B^z = -\p_{\mu}^2\chi = \frac{\rho_s}{\rho_s + {\bf p}^2/4\pi}\tilde{n}_{S} (p), 
 \eea
 so that the total flux is equal to the topological charge of the ${\bf n}$-field.
 
 Substituting (\ref{sol}) into (\ref{gaussian}), we obtain
\begin{eqnarray}\label{l}
G_{\eta \chi }& =&
 \frac{1}{2}{\int  \frac{d^2 p}{(2\pi)^{2}}
}\left[
\rho_s {\bf p}^{2} 
\frac{|\tilde{n}_{S} (p)|^{2}}{( 4\pi\rho_s + {\bf p}^2)^{2}}
+\frac{1}{4\pi}\frac{|\rho_{s}\tilde{n}_{S} (p)|^{2}}{(\rho_{s}+ {\bf p}^{2}/4\pi)^{2}}
\right]\cr
&=&
 \frac{1
}{8\pi}
{ \int  \frac{d^2 p}{(2\pi)^{2}}}
\frac{\rho_{s}|\tilde{n}_{S} (p)|^{2}}{(\rho_{s}+ {\bf p}^{2}/4\pi)^{2}}
\left[\frac{{\bf p}^{2}}{4\pi}+\rho_{s}\right]\cr
&=& \frac{1}{8\pi}
{\int  \frac{d^2 p}{(2\pi)^{2}}}
\frac{|\tilde{n}_{S} (p)|^{2}}{(1 +{\bf p}^{2}\lambda_{L}^{2})},
\end{eqnarray}
where $\lambda_{L}^{2} = 
\frac{1}{4\pi \rho_{s}}
\equiv \frac{1}{4\pi \rho_{s}q^{2}}
$, reinstating $q$.
Adding back the gradient term $\frac{\rho_{\perp }}{2}
(\partial_{\mu}\bn )^{2}$, and converting back to real-space, we obtain
 \bea
 &&  G = 
\int d^{2}x \frac{\rho_{\perp}}{2}(\p_{\mu}{\bf n})^2 + \frac{1}{8\pi}
\int d^{2}x d^{2 }x'\tilde{n}_{S}(x)V(|\bx-\bx'|)\tilde{n}_{S}(\bx'), \nonumber\\
 &&  V(p) = \frac{1}{1+ {\bf p}^{2}\lambda_{L}^{2}}. \label{free2}
 \eea
Now if we reinstate the external field $H= H_{z}$, we note that
\begin{eqnarray}\label{l}
&&- \int d^{2}x\frac{1}{4 \pi} 
HB (x)\notag\\
& =& - \frac{1}{4 \pi} H \int d^2 x  \int 
{\frac{d^{2}p}{(2\pi)^{2}}}
 B(p) e^{i {\bf x} \cdot {\bf p}}\cr
&=& - \frac{1}{4 \pi}  H \int d^2 x \int {\frac{d^{2}p}{(2\pi)^{2}}}
 V (p)\tilde{n}_{S} (p)e^{i {\bf x} \cdot {\bf p}}\cr
&=&- \frac{1}{4 \pi} H \int d^{2}x 
d^{2}x' d^2x''
V (x'')\tilde{n}_{S} (x')  \notag\\
&&\times {  {\int \frac{d^{2}p}{(2\pi)^{2}}}
} e^{i {\bf p} \cdot ({\bf x-x'-x''})},\cr
&=&- \frac{1}{4\pi} H \int d^{2}x  d^{2}x' 
V (|x-x'|)\tilde{n}_{S} (x'),\cr
&&
\end{eqnarray}
where the normalization factor of the Fourier transformation is
$F(x)=\frac{1}{(2\pi)^2}\int d^2 p F(p)e^{i {\bf p} \cdot {\bf x}}$,
$F(p)=\int d^2 x F(x)e^{-i {\bf p} \cdot {\bf x}}$.
We can then incorporate the $-{\bf H}\cdot {\bf B}$ term by writing 
 \bea
&G =& 
\int d^{2}x 
\left[\frac{\rho_{\perp}}{2}(\p_{\mu}{\bf n})^2 
 - \frac{H^{2}}{8\pi}
 \right]  \notag\\
 &&+ \frac{1}{8\pi}
\int d^{2}x d^{2 }x' \delta \tilde{n}_{S} (\bx )
V(|\bx-\bx'|)\delta \tilde{n}_{S} (\bx' ),
\cr 
&  V(p) =& \frac{1}{1+ {\bf p}^{2}\lambda_{L}^{2}}. 
\label{free4}
\label{free3}
 \eea
where $\delta {\tilde{n}_{S}} (x)=\tilde{n}_{S}(x)-H \equiv  \Phi_{0}n_{S} (x)-H$.

\section{The microscopic Hamiltonian---topological CMT model}

The microscopic Hamiltonian is motivated by a Majorana representation of local moments first 
developed by Coleman, Miranda and Tsvelik (CMT model)\cite{Coleman94, Coleman_PhysicaB1993}. 
In order to incorporate the topological aspects of SmB$_6$, we consider a `p-wave' Kondo-Heisenberg model.

\begin{eqnarray}
H&=&\sum_{k,\sigma}(\epsilon_{{\bf k}} -\mu)c_{{\bf k},\sigma}^\dagger c_{{\bf k},\sigma} 
-J_K\sum_{j, \alpha, \beta} (\tilde{c}_{j\alpha}^\dagger \boldsymbol{\sigma}_{\alpha \beta}
\tilde{c}_{j\beta})\cdot {\bf S}_{j} \notag\\
 &&- J_H \sum_{\langle i,j \rangle} {\bf S}_i \cdot {\bf S}_j,
\label{Hamiltonian}
\end{eqnarray}
where 
 $c_{{\bf k} \sigma}$ is the electron operator,  $\epsilon_{{\bf k}} $ and $\mu$ are the dispersion of 
 conduction electrons and  chemical potential, ${\bf S}$ is the Sm$^{+3}$ local moment. $J_K$ and $J_H$ are the Kondo and 
 Heisenberg exchange. 
 $\tilde{c}_{i\alpha}$ represents the local Wannier orbital which is defined as $\tilde{c}_{i\alpha} = \sum_{j}\Phi_{i,j}^{\alpha, \beta} c_{j \beta}$ in terms of the original conduction electrons.
 Here, $\Phi^{\alpha, \beta}_{i,j}$ is a `p-wave' form factor, $\Phi_{i,j}^{\alpha, \beta}=-\frac{i}{2} \hat{r}_{ij} \cdot \boldsymbol{\sigma}_{\alpha,\beta}$ 
 which results from net angular momentum difference $|\Delta l | = 1$ 
 between the heavy $f$ and light $d$ electrons\cite{Alexandrov_PRL2015}. 
 Next we use a Majorana representation for the spin
 \begin{eqnarray}
  {\bf S}_{j} \to -\frac{i}{2} \boldsymbol{\eta}_{j} \times \boldsymbol{\eta}_{j}
  \label{Majorana}
 \end{eqnarray}
 where $\boldsymbol{\eta}_{j}\equiv (\eta^{1}_{j},\eta^{2}_{j},\eta^{3}_{j})$ 
is a three component vector of Majorana
 fermions, defined at each site $j$. This representation of the spin ensures that
 the spin commutation relations and $S^2=3/4$ are satisfied. Also note that in this representation the constraint
 field is automatically fulfilled. Inserting Eq. (\ref{Majorana}) into Eq. (\ref{Hamiltonian})
\begin{eqnarray}
H&=&\sum_{k,\sigma}(\epsilon_{{\bf k}} -\mu)c_{{\bf k},\sigma}^\dagger c_{{\bf k},\sigma} 
-J_K\sum_{\substack{j,  \alpha,\beta}} \tilde{c}_{j\alpha}^\dagger \boldsymbol ( \boldsymbol \sigma_{\alpha\beta} \cdot {\boldsymbol \eta}_j)^2\tilde{c}_{j \beta}  \notag\\
&&-\frac{J_{H}}{2}\sum_{\langle i,j \rangle} ( {-i \boldsymbol \eta_i \cdot \boldsymbol \eta_j})^2
\end{eqnarray}
We have used the identity $i \boldsymbol{\sigma} \cdot( \boldsymbol{\eta}_j \times \boldsymbol{\eta}_j )= (\boldsymbol{\sigma}\cdot\boldsymbol{\eta}_j)^2-3/2$. 
Next we carry out a mean field decoupling by a Hubbard-Stratonovich
transformation on the Kondo interaction: 
\begin{eqnarray}
&&-J_{K}\biggl[ \tilde{c}^{\dagger}_{j\alpha} (\boldsymbol{\sigma}_{\alpha \gamma} \cdot \boldsymbol{\eta}_j) \biggl] 
\biggl[ ( \boldsymbol{\sigma}_{\gamma \beta} \cdot \boldsymbol{\eta}_j)\tilde{c}_{j\beta} \biggl]  \notag\\
\to &&\biggl[\tilde{c}_{j\alpha}^\dagger (\boldsymbol{\sigma}_{\alpha \beta} \cdot
{\boldsymbol \eta_j}) {V}_{j\beta} +{\rm h.c}\biggr] + 
\frac{ {V}_{j\gamma}\dg {V}_{j\gamma}}{J_{K}},
\label{Eq: HS}
\end{eqnarray}
where the auxiliary field ${V}_{i\beta}$ is 
a fluctuating variables, integrated within a path integral.
In the mean field treatment, we assume ${V}_{j\beta}$ assumes a fixed
value at each site. The last term in Eq. (\ref{Eq: HS}) is only
important in evaluating the mean-field equations, and will be dropped
in the discussion that follows. 

We also perform a Hubbard-Stratonovich transformation on  the
spin-fluid, 
\begin{equation}\label{}
-\frac{J_{H}}{2}\sum_{\langle i,j \rangle}(-i{\boldsymbol \eta}_{i} \cdot {\boldsymbol \eta}_{j})^2 
\to \sum_{\langle i,j \rangle} (-i  \Delta_{ij} {\boldsymbol \eta}_{i} \cdot {\boldsymbol \eta}_{j})  + \frac{\Delta_{ij}^2}{2J_H}
\end{equation}
where the field $\Delta_{ij}$ is a real, odd-function of position,
$\Delta_{ij}= -\Delta_{ji}= \Delta_{ij}^{*}$. In the simplest
mean-field theory, taking the bonds to be uniform, 
$\Delta_{i, i+\hat {\bf a}}= \Delta $ ($\hat {\bf a}= (\hat x,\hat y,
\hat z)$), this term leads to 
to a momentum-dependent dispersion of Majorana Fermions as follows
\begin{equation}\label{}
\sum_{\langle i,j \rangle} -i \Delta_{ij} {\boldsymbol \eta}_{i}
\cdot {\boldsymbol \eta}_{j} = \sum_{\bk \in \frac{1}{2}BZ}\epsilon_f({\bf k}) {\boldsymbol
\eta_{\bf k}}^\dagger\cdot {\boldsymbol \eta_{\bf k}},
\end{equation}
where
$\epsilon (\bk )= 2 \Delta (\sin k_{x}+\sin k_{y}+ \sin k_{z})$. Here,
the
momentum sum is restricted over half the Brillouin zone ($\bk \in
\frac{1}{2}BZ$), 
because  the Fourier transform of real Majorana operators satisfies
$
{\boldsymbol
\eta }_{-\bk }= 
{\boldsymbol
\eta }\dg _{\bk }= { \frac{1}{\sqrt{N}}\sum_{i} e^{i {\bf k}\cdot {{\bf R}}_i}}\boldsymbol{\eta}_i^\dagger
$ with $\boldsymbol{\eta}_i^\dagger=\boldsymbol{\eta}_i$, so that the creation and annihilation operators of the Majorana
fields are only independent in half the Brillouin zone. 
In the following discussion, we will neglect the constant term $ \Delta_{ij}^2/2J_H$.

In the original CMT model\cite{Coleman94, Coleman_PhysicaB1993},
a ``staggered'' order parameter (${\cal V}_j= \exp (i \bQ \cdot
\bR_j/2){\cal V}_{0} $ 
was found to have the lowest mean-field free energy, where $\bQ  =
(\pi,\pi,\pi)$ defines a staggered wavevector. In momentum space, this
has the effect of transfering a momentum $\bQ /2$ each time a Majorana
fermion converts into an electron. The Fourier transformed 
hybridization term is written
\begin{eqnarray}\label{hyb1}
H_{hyb} = \sum_{\bk  } \left[\tilde{ c}\dg _{\bk } 
(\boldsymbol{\sigma }\cdot
\boldsymbol{\eta }_{\bk -\bQ /2})
 V_{0}+ V_{0}\dg (\boldsymbol{\sigma }\cdot
\boldsymbol{\eta }\dg _{\bk -\bQ /2}) \tilde{ c}_{\bk }
\right]  \notag\\
\end{eqnarray}
\begin{widetext}
Now by taking the transpose of the second term, we can rewrite the 
hybridization term in the alternate form 
\begin{align}\label{hyb2}
H_{hyb} 
&= \sum_{\bk  } \left[\tilde{ c}^{T} _{\bk } (-i \sigma_{2}) 
({-}\boldsymbol{\sigma }
\cdot
\boldsymbol{\eta }\dg _{\bk -\bQ /2})
(i\sigma_{2}) V^{*}_{0}+ h.c 
\right]
&
(-\boldsymbol{\sigma }^{T}= -i\sigma_{2}\boldsymbol{\sigma
}i\sigma_{2})
\cr
&= \sum_{\bk  } \left[\tilde{ c}^{T} _{\bk } (-i \sigma_{2})
(\boldsymbol{\sigma }\cdot
\boldsymbol{\eta } _{-\bk +\bQ /2})
(i\sigma_{2}) V^{*}_{0}+ h.c 
\right]
&
(-\boldsymbol{\eta}\dg_{\bk -\bQ /2}= 
\boldsymbol{\eta}_{-\bk +\bQ /2})
\cr
&= \sum_{\bk  } \left[\tilde{ c}^{T} _{-\bk } (-i \sigma_{2})
(\boldsymbol{\sigma }\cdot
\boldsymbol{\eta } _{\bk +\bQ /2})
(i\sigma_{2}) V^{*}_{0}+ h.c 
\right]&(\bk \rightarrow -\bk )
\cr
&= \sum_{\bk  } \left[\tilde{ c}^{T} _{-\bk+\bQ  } (-i \sigma_{2})
(\boldsymbol{\sigma }\cdot\boldsymbol{\eta } _{\bk -\bQ /2})
(i\sigma_{2}) V^{*}_{0}+ h.c 
\right]&(\bk \rightarrow \bk { -} \bQ )\cr
&=\sum_{\bk  } \left[{ c}^{T} _{-\bk+\bQ  } (i \sigma_{2})
(\sin k_{l}\ \sigma_{l}) 
(\boldsymbol{\sigma }\cdot\boldsymbol{\eta } _{\bk -\bQ /2})
(i\sigma_{2}) V^{*}_{0}
+ h.c \right]
&(\tilde{c}^{T}_{\bk } (-i\sigma_{2}) = 
c^{T}_{\bk } (i\sigma_{2}) (\sin k_{l}\ \sigma_{l}) )
\end{align}
\end{widetext}
Combining (\ref{hyb1}) and (\ref{hyb2} ), we can write the
hybridization in the form 
\begin{eqnarray}\label{l}
H_{hyb} &=& \frac{1}{2}
\sum_{\bk }\chi \dg_{\bk } \left[
(\sin
k_{l}\sigma^{l})(\boldsymbol{\sigma }\cdot\boldsymbol{\eta } _{\bk
-\bQ /2}){\cal V}_{0}+ h.c
 \right]\cr
&=& 
\sum_{\bk \in \frac{1}{2}BZ}\chi \dg_{\bk } \left[
(\sin
k_{l}\sigma^{l})(\boldsymbol{\sigma }\cdot\boldsymbol{\eta } _{\bk
-\bQ /2}){\cal V}_{0}+ h.c
 \right],  \notag\\
\end{eqnarray}
where $\sin k_{l}\sigma^{l}\equiv \sin k_{x}\sigma_{x}+ \sin
k_{y}\sigma_{y} + \sin k_{z}\sigma_{z}$,  while
\begin{equation}\label{}
\chi_{\bk } = \pmat{c_{\bk}\cr
-i \sigma_{2}c\dg _{-\bk+ \bQ }} 
\end{equation}
and 
\begin{equation}\label{}
{\cal V}_{0} = \pmat{V_{0}\cr i \sigma_{2}V^{*}_{0}}.
\end{equation}
are four component Balian-Werthammer spinors (the minus
sign-distinction is deliberate).  Note that the sum over
half the Brillouin zone ensures that all creation and annihilation 
operators in the Hamiltonian are independent. 

\begin{widetext}
In this same notation, the conduction electron Hamiltonian is given by 
\begin{align}
H_c=\sum_{\bf k}\chi_{\bf k}^\dagger 
\left(\begin{array}{cccc}
\epsilon_{\bf k}-\mu & 0 & 0 & 0 \\
0 & \epsilon_{\bf k}-\mu & 0 & 0 \\
0 & 0 & -\epsilon_{-{\bf k}+ \bQ } +\mu & 0 \\
0 & 0 & 0 & -\epsilon_{-{\bf k}+ \bQ } +\mu\end{array}\right)
 \chi_{\bf k}.
\end{align}
\end{widetext}

In order to simplify our calculation, we consider the particle-hole symmetric case
\begin{align}
\epsilon_{{\bf k}}&=-\epsilon_{-{\bf k}+ \bQ }\equiv \epsilon_c(\bf k)  \notag\\
&=-2t_{c} (\cos k_x +\cos k_y+\cos k_z)+8t'_{c} \cos k_x\cos k_y\cos k_z.
\end{align}
We use Pauli matrices $\boldsymbol{\tau}$ as the particle-hole basis in the Balian-Werthamer four-spinor notation.
The conduction electron Hamiltonian has a compact form
\begin{align}
H_c=\sum_{\bf k\in \frac{1}{2}BZ}\chi_{\bf k}^\dagger (\epsilon_c(\bf k) -\mu \tau_3) \chi_{\bf k}.
\end{align}

Finally, the mean field Hamiltonian can be written as, $H=\sum_{\bf k} \psi_{\bf k}^\dagger \mathcal{H}({\bf k})
 \psi_{\bf k}$ with the single-particle Hamiltonian 
\begin{align}
\mathcal{H}({\bf k})=\left(\begin{array}{cc}\epsilon_c({\bf k}) - \mu \tau_3 & {\cal V}({\bf k}) \\{\cal V}({\bf k})^\dagger & \epsilon_f ({\bf k})\end{array}\right),
\end{align}
where $\psi_{\bf k} = ({c}_{{\bf k}\uparrow}, {c}_{{\bf
k}\downarrow},- {c}^\dagger_{-{\bf k}+\bQ \downarrow }, {c}^\dagger_{-{\bf
k}+\bQ \uparrow},\eta_{{\bk-\bQ /2}}^x,$
$\eta_{{\bk- \bQ /2}}^y,\eta_{{\bk -\bQ /2}}^z) ^{\rm
T}$ and 
\begin{equation}\label{}
{\cal V} (\bk )_{\alpha }^{\ \ a} = (s_{k_{l}} \ \sigma^{l})_{\alpha \beta }\sigma^{a}_{\beta \gamma} ({\cal V}_{0})_{\gamma}.
\end{equation}
where we use summation convention and 
the short-hand $s_{k_{l}}\equiv  \sin k_{l}$.
If we choose ${\cal V}_{0} = (V,0,0,V)^{T}$, then 
the hybridization $V({\bf k})$ takes the form
\begin{align}
{\cal V}({\bf k})=V\left(\begin{array}{ccc} s_{k_x} - i s_{k_y} & s_{k_y} + i s_{k_x} & s_{k_z} \\
-s_{k_z} & - i s_{k_z} & i s_{k_y} +s_{k_x} \\
-s_{k_z} & i s_{k_z} & -i s_{k_y} + s_{k_x} \\
-s_{k_x} - i s_{k_y} & -s_{k_y} + i s_{k_x} & -s_{k_z}\end{array}\right).
\end{align}

 \subsection{Bulk and surface spectra}
 
The bulk spectrum of the single-particle Hamiltonian has
a gapless band shown in Fig. \ref{Fig1_old}(a) .
In particular, when $\mu=0$,
the gapless quasiparticle lead to a form
\begin{eqnarray}
\phi_{\bf k} &=& \frac{1}{\sqrt{2(s_{k_x}^2+s_{k_y}^2+s_{k_z}^2)}} \Big[ s_{k_z} {c}_{{\bf k}\uparrow} +(s_{k_x}-is_{k_y}){c}_{{\bf k}\downarrow}  \notag  \\
&&+(s_{k_x}+is_{k_y}){c}^\dagger_{-{\bf k+Q} \downarrow}+s_{k_z} {c}^\dagger_{-{\bf k+Q}\uparrow} \Big],
\end{eqnarray}
where $\phi_{\bf k} = \phi^\dagger_{-{\bf k+Q}}$ satisfies the "twisted Majorana condition".

In the slab geometry,
we observe surface states on $(100)$ and $(010)$ surfaces.
The crossings in the $(100)$ surface spectrum are located at $X_1=(k_y, k_z)=(\pi,0)$
and $X_2=(k_y, k_z)=(0,\pi)$[see Fig. \ref{Fig1_old}(b)(c)].
We find the surface states are robust under mirror symmetry preserving perturbations.

\begin{widetext}

The stability analysis of the robustness of the surface states is done by adding various symmetry preserving perturbation.
Specifically, nonvanishing chemical potential $\mu$ and 
Zeeman terms on the conduction electrons,
$+ \mu B_z {c}^{\dagger}_{\uparrow}{c}_{\uparrow}$ and 
$- \mu B_z {c}^{\dagger}_{\downarrow}{c}_{\downarrow}$,
do not break the mirror symmetry $\mathcal{R}_z$.
We observe the crossings in the $(100)$ and $(010)$ surface spectra
cannot be gapped by adding these mirror symmetry-preserving perturbations.
We now analyze the topological origin of the mirror symmetry-protected surface 
states using an entanglement spectrum and Berry phase analysis.

\begin{figure}[htbp]
\centering
\includegraphics[height=9 cm] {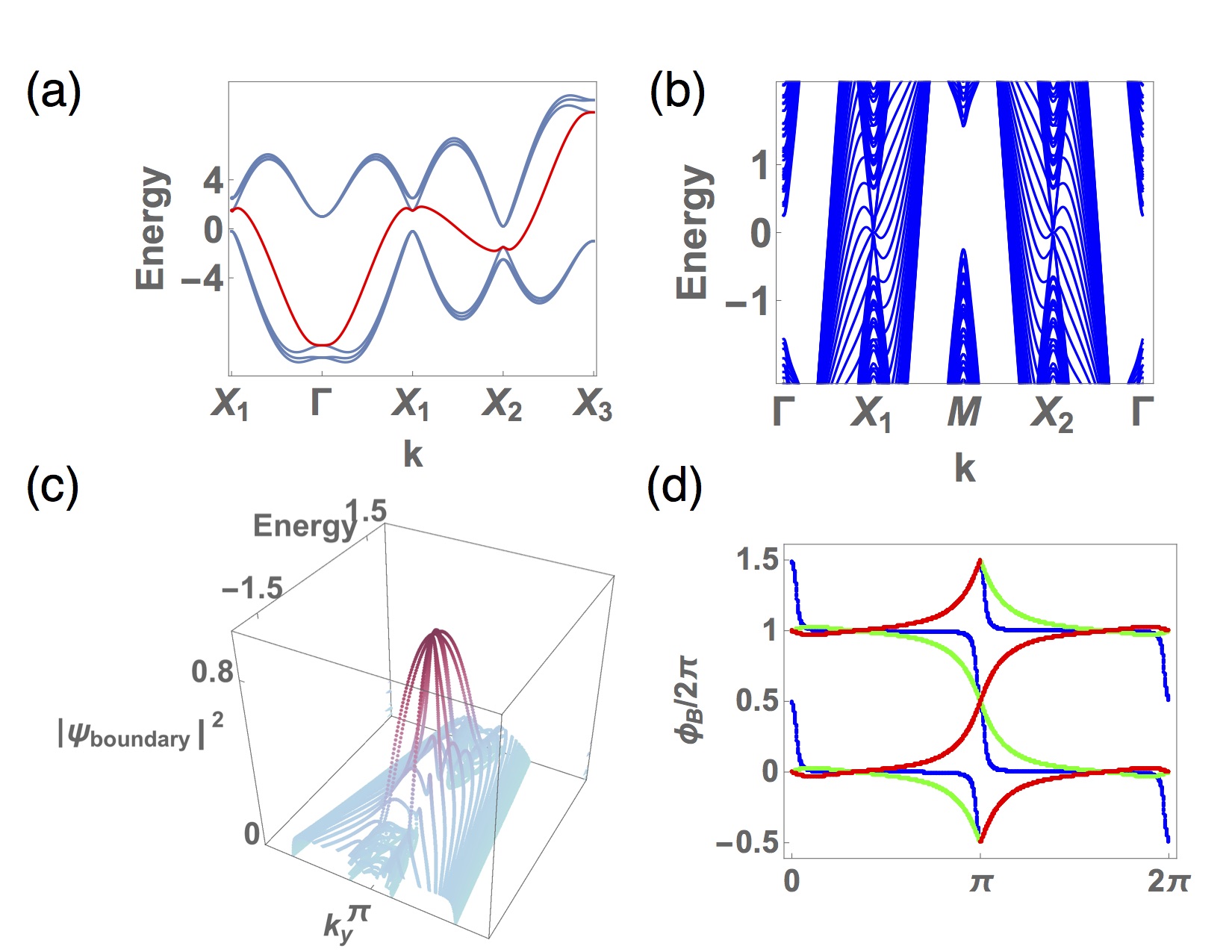}
\caption{(a) The bulk spectrum with $X_1=(\pi,0,0)$, $X_2=(\pi,\pi,0)$, and $X_3=(\pi,\pi,\pi)$. The red line indicates the gapless Majorana band in the CMT model.
(b) The $(100)$ surface spectrum with $X_1=(k_y=\pi,k_z=0)$, $M=(\pi,\pi)$, and $X_2=(0,\pi)$. The crossings
in the $X_1$ and $X_2$ are surface states, which the density profile in shown in (c).
(c) The density of surface wave function $|\psi_{\rm boundary}|^2$ and energy as a function of $k_y$ at $k_z=0$.
The crossings at $k_y=0$ are localized states on the $(100)$ surface.
(d) The Berry phase as a function of $k_y$ at $k_z=0$ for the lowest three occupier bands (Blue, Red, Green)
with corresponding mirror eigenvalue $(++-)$. The parameters in the CMT model are
$(t_c, t'_c, \alpha, \mu, V)=(1, 0.5,  0.1, 0.5,5)$}
\label{Fig1_old}
\end{figure}
\end{widetext}

\subsection{Mirror symmetry protected boundary states and its topological origin}

The nonvanishing order parameter $\mathcal{V}$ breaks the
 time-reversal symmetry and spin-rotation symmetry in the system. 
 The remaining symmetries are the inversion symmetry and the reflection symmetry $\mathcal{R}_z$
in the CMT model with $\mathcal{P}^\dagger \mathcal{H}({\bf k})\mathcal{P}=\mathcal{H}(-{\bf k})$
and $\mathcal{R}_z^\dagger \mathcal{H}(k_x,k_y,k_z)\mathcal{R}_z=\mathcal{H}(k_x,k_y, -k_z)$, respectively.
The matrix representation of the inversion symmetry and reflection symmetry are
\begin{align}
\mathcal{P}=\left(\begin{array}{cc}\tau_0 \sigma_0 & 0 \\0 & -\mathbb{I}_{3 \times 3}\end{array}\right), \quad
\mathcal{R}_{z}= \pmat{ \tau_{3}\sigma_{3} &  \cr & R_{z}},
\end{align}
where $R_{z}$ is the three dimensional reflection matrix,
\begin{equation}\label{}
R_{z}=\pmat{1 & 0 & 0\\ 0 & 1& 0 \\ 0 & 0 &-1}.
\end{equation}

One should be noticed that 
the gapless Majorana state is orthogonal to the remaining six gapped states in the energy basis.
Thus, the Hilbert space of the CMT model is the direct sum of the two sub-Hilbert spaces, 
$\mathbb{H}_{\rm CMT}=\mathbb{H}_{\rm gapped} \oplus \mathbb{H}_{\rm gapless}$, where $\mathbb{H}_{\rm gapped/gapless}$
are the sub-Hilbert spaces of the gapped/gapless states respectively.
The mirror symmetry protected surface states originate from the gapped bands
and we need to extract the information from the gapped sub-Hilbert space $\mathbb{H}_{\rm gapped}$.
In order to do this, we perform an entanglement spectrum
analysis\cite{Hughes11, Chang14} and a Berry phase analysis
\cite{Taherinejad14} for the three lowest occupied bands which belong to $\mathbb{H}_{\rm gapped}$.

\begin{widetext}

\begin{figure}[htbp]
\centering
\includegraphics[height=4.5 cm] {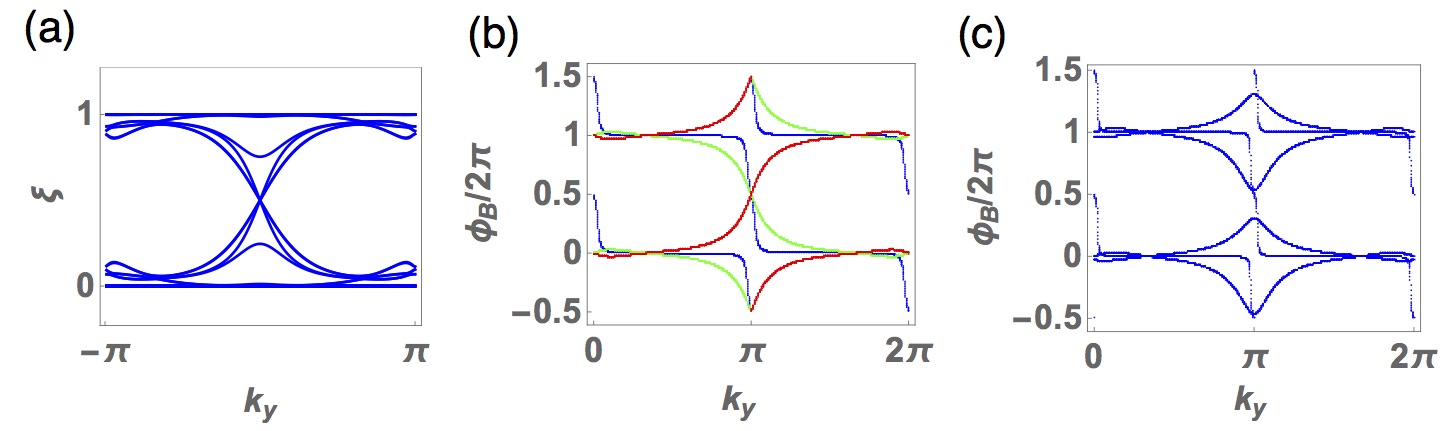}
\caption{(a) Entanglement spectrum $\xi(k_y)$ at $k_x=\pi$ for the bipartition along $z$ direction. 
(a)The Berry phase as a function of $k_y$ at $k_z=0$ for the lowest three occupier bands (Blue, Red, Green)
with corresponding mirror eigenvalue $(++-)$. (b) The crossing between blue and red lines can be gapped by adding inversion breaking and $\mathcal{R}_z$ preserving term.}
\label{Berry}
\end{figure}

\end{widetext}

In the entanglement spectrum, there are robust mid-gap states protected by inversion symmetry.
In Fig. \ref{Berry}(a), the entanglement Hamiltonian is constructed by a bipartition along $z$ direction
and the entanglement spectrum is plotted as a function of $k_y$ at $k_x=\pi$.
We observe there are six mid-gap states at $k_y=0$ which are protected by inversion symmetry. 
The topological invariant of inversion symmetric topological
phases is given by the number of mid-gap states in the entanglement
spectrum, which can be computed from counting the inversion eigenvalues
of the occupied bands at inversion symmetric points
\begin{align}
\nu({\bf k}_\perp^0)=2|N({\bf k}_\perp^0, k_\parallel=0)-N({\bf k}_\perp^0, k_\parallel=\pi)|,
\end{align}
where ${\bf k}_\perp^0$ is the inversion symmetric points for the momenta perpendicular 
to the bipartition direction.  $N({\bf k}_\perp^0, k_\parallel)$ is
the number of negative inversion eigenvalue of the occupied band
at $k_\parallel=0$ or $k_\parallel=\pi$,  with $k_\parallel$ being the momentum along the bipartition direction\cite{Hughes11}. 
For example, if the bipartition is along $z$ direction, $ k_\parallel=k_z$, and ${\bf k}_\perp^0=(k_x,k_y)=(0,0)$, $(0,\pi)$,
$(\pi, 0)$, and $(\pi,\pi)$. 

In the CMT model, the number of mid-gap states is six at both ${\bf k}_\perp=(0,\pi)$ and $(\pi,0)$
for the bipartition direction along $x$, $y$ and $z$ directions.
Although there is no inversion-symmetry protected surface state in the physical surface spectrum,
the mid-gap states in the entanglement spectrum cannot be removed
and cannot be adiabatically connected to a system without mid-gap states\cite{Hughes11, Chang14}.
Thus the CMT model can be seen as an inversion-symmetric topological phase.

On the other hand, we observe physical surface states on $(100)$ and $(010)$
surfaces. These surface states are protected by
the reflection symmetry $\mathcal{R}_z$. 
The topological invariant of the mirror-symmetric topological phases
is the mirror Chern number at the mirror planes ($k_z=0$ and $\pi$).
The Chern number is defined as
\begin{align} 
n = \int d {\bf a}\cdot  \nabla \times {\bf A} = \oint d {\bf l }\cdot {\bf A},
\label{Eq: Chern}
\end{align}
where $\int d {\bf a}$ is the embedded Brillouin zone of the mirror plane,
$\oint d {\bf l}$ is the loop integral along the boundary of the
embedded Brillouin zone of the mirror plane,
and $A_i ({\bf k})= \langle u  ({\bf k})|\partial_{k_i} u ({\bf k}) \rangle $ is the Berry connection with $|u  ({\bf k})\rangle$ being the occupied band.
We can compute the Chern number by introducing the Berry phase $\phi_B(k_{\bar{i}})=\int^{2\pi}_0 d{k_i} A_i(k_i,k_{\bar{i}})/(2\pi)$
with momenta ${\bf k}=(k_i,k_{\bar{i}})$ in the mirror plane.
For example, at the mirror plane momenta 
$k_z=0, \pi$, we first compute the Berry phase $\phi_B(k_y)=\int^{2 \pi}_0 d{k_x} A_x(k_x,k_y)/(2\pi)$ and monitor the 
Berry phase winding around $k_y  \to k_y+2 \pi$. 
The Chern number in Eq. (\ref{Eq: Chern}) is defined by a loop
integral, which  can
rewritten as the difference of the Berry phase at $k_y+2\pi$ and $k_y$. 
Thus the Chern number is $n=\phi_B(k_y)-\phi_B(k_y+2\pi)$.
As shown in Fig. \ref{Berry}(b), the Berry phase for the lowest three occupied bands (Red, Green, Blue)
with corresponding mirror eigenvalues $(++-)$, gives the corresponding Chern number $(+1,-1,-2)$ [At $k_z=0$].
Hence the mirror Chern number is $n_M=|n_+-n_-|=2$ with $n_{\pm}$ being the Chern number of the bands with $\pm$ mirror eigenvalue.
The $"+"$ mirror eigensector has zero Chern number which indicates the crossing of the Berry phase flow 
between two states with $"+"$ mirror eigenvalue can be gapped [red and green lines in Fig. \ref{Berry}(b)].
This crossing is due to the inversion symmetry and can be removed by
introducing a inversion breaking 
and $\mathcal{R}_z$ preserving term [Fig. \ref{Berry}(c)].
Notice that the Berry phase spectrum mimics 
the surface spectrum. The three spectral flows in the Berry phase spectrum
indicates there are two chiral surface states on $(100)$ surface witch agrees with our numerical observation.

 \subsection{Quantum oscillations}
In order to determine whether a Majorana Fermi surface exhibit quantum oscillations, we calculate 
the dispersion under magnetic field by projecting the Hamiltonian on the Majorana band.
\begin{align}
\epsilon^{\rm M}_{\bf k, A}=\langle \phi^{\rm M}_{\bf k}|H(k,A)|\phi^{\rm M}_{\bf k}\rangle =\frac{1}{2}(\epsilon^{\rm e}_{{\bf k}-e{\bf A}}+\epsilon^{\rm h}_{{\bf k}+e{\bf A}}),
\end{align}
where $\epsilon^{\rm e}_{\bf k-eA}$ and $\epsilon^{\rm h}_{\bf k+eA}$
are the dispersion for electrons and holes bands that couple to the
external gauge field with opposite signs. Although $\epsilon^{\rm M}_{\bf k, A}$ break 
gauge invariance,we checked that our results are independent of the gauge choice for spherical Fermi 
surfaces. A fully gauge invariant calculation under magnetic field require a background of skyrmion fluid. 
In the main text, Fig. 2(c), we show that the density of states arising from the Majorana band under magnetic field indeed have 
sharp and periodic features that resemble Landau levels. These calculations are done in a two dimensional system with the Landau 
gauge ${\bf A} = (0, Bx, 0)$, giving rise to a perpendicular magnetic field ${\bf B} = \nabla \times {\bf A} = B\hat{z}$. 
Since gauge field breaks the translational invariance along $x$ direction, we inverse Fourier 
transform the Hamiltonian back to complex fermions and perform a real space calculation 
along $x$ direction and include the magnetic field using the Peierls substitution, 
$t_{ij} \rightarrow t_{ij} {\rm e}^{\int_{i}^j {\bf A}\cdot d{\bf l}}$. Since $k_y$ is still a good 
quantum number, the calculation reduces to one dimensional 
strip for a given $k_y$, which are then summed up within the one dimensional Brillouin zone.
To gain further insight, we can consider
a parabolic band for the dispersion
\begin{eqnarray}
\epsilon^{\rm M}_{\bf k, A}&\simeq& \frac{1}{2}\Big[ \frac{({\bf k}-e{\bf A})^2}{2m^*}+\frac{({\bf k}+e{\bf A})^2}{2m^*}\Big] \nonumber \\
&=&\frac{|{\bf k}|^2+e^2|{\bf A}|^2}{2m^*}
\end{eqnarray}
Since the linear coupling term ${\bf k}\cdot {\bf A}$ cancels, the current operator $J_\alpha = \partial \epsilon^{\rm M}_{\bf k, A} /\partial
A_\alpha|_{\bf A=0}=0$ vanishes. Nevertheless, the $A_\alpha^2$ term, responsible for Landau quantization is still present, 
which is responsible for the Landau level like features in the density of states. Since quantum oscillations 
are manifestations of the sharp and periodic features of the Landau levels, we anticipate that Majorana Fermi surface can 
also give rise to quantum oscillations.  

\section{Robust Order-parameter isotropy  in the presence of spin-orbit coupling.}\label{isotropic}

One thing that might save the superconductor, is if it has a spin
anisotropy, induced for example, by spin-orbit coupling.  If such an
anisotropy set up an easy plane for the vector $\bn$, then the
system would reduce to a $U (1)$ order parameter, with stable
vortices. 
However, the
Free energy of the ground-state does not depend on the orientation of
the order-parameter. This can be seen by integrating out the
conduction electrons, writing the effective action (Free energy) as
\[
F = - \frac{T}{2}\sum_{\bk, i\omega_{n} }{\rm Tr}\ln 
\left[-{\cal  G}^{-1}[\bk ,i\omega_{n}] \right]
\]
where 
\[
{\cal G}^{-1}[\bk ,i\omega_{n}]= ( i\omega_{n}- \epsilon_{f} (\bk
))\mathbb{I}_{3 \times 3} - \Sigma_{M} (\bk ,i\omega_n)
\]

Now the only dependence of $F$ on the order parameter orientation
comes via the Majorana self energy $\Sigma_{M}$. In our model, this quantity is
given by 
\begin{eqnarray}\label{mayse}
\Sigma^{M}_{ab} (\bk ,\omega) &=& {\cal V}\dg 
\sigma^{a}({\bf s}_{\bk }\cdot \vec{\sigma }) 
\frac{1}{i\omega_{n} - \epsilon_{\bk}+\mu\tau_{3}}
({\bf s}_{\bk }\cdot
\vec{\sigma })\sigma^{b}  {\cal V}  \cr
&=&  s_{\bk }^{2}
{\cal V}\dg 
\frac{\sigma^{a}\sigma^{b}
}{i\omega_{n} - \epsilon_{\bk}+\mu\tau_{3}}{\cal V}
\cr &=& 
\frac{s_{\bk }^{2}V^{2} }{(i\omega_{n}-\epsilon_{\bk })^{2}- \mu^{2}}
\left[\delta^{ab} (i\omega_{n}-\epsilon_{\bk })
 - \mu i \epsilon^{abc}\hat n_{c}
\right]\cr &=&
\Sigma_{0} (\bk,i\omega_{n})\delta^{ab} - i \Sigma_{1} (\bk,i\omega_{n}) \epsilon^{abc}n_{c}
\end{eqnarray}
where ${\cal V}\dg {\cal V}= V^{2}$, ${\cal V}\dg \tau_{3}\vec{ \sigma
}{\cal V} = V^{2 }\bn$. 
Notice how this self energy has  complete mirror symmetry ($
\Sigma_{M}(k_{x},k_{y},k_{z},i\omega_{n})
=\Sigma_{M}(\pm k_{x},\pm k_{y},\pm k_{z},i\omega_{n})
$), but it also contains an anisotropy dependent on the direction
$\bn$.  However, this does not produce any corresponding anisotropy in the
free energy.  To see this, lets look at the Free energy of the Majorana Fermions
\begin{widetext}
\begin{align}
\label{l}
F =- \frac{T}{2} \sum_{\bk, i\omega_{n} }{\rm Tr}\ln  
[ (-i\omega_{n}+ \epsilon_{f} (\bk ) +
\Sigma_{0})\delta^{ab}  
+  \Sigma_{1} (\bk ,i\omega_{n})
 i \epsilon^{abc}n_{c}  ].
\end{align}

Now we can always
carry out a rotation in spin space inside the trace 
so that the
n-vector points in the z-direction, and this rotation leaves the trace
and hence the energy, unchanged. Since the Free energy that results
has lost all information about the direction of the n-vector, it
follows that the Free energy is isotropic. Lets see this in its
engineering glory:

\begin{eqnarray}\label{l}
F &=& - \frac{T}{2}\sum_{\bk, i\omega_{n} }{\rm Tr}\ln 
\left[ \pmat{
\epsilon_{fk}+ \Sigma_{0}-i\omega_{n} 
&0 &0 \cr
0 & \epsilon_{fk}+ \Sigma_{0}-i\omega_{n} 
& - i \Sigma_{1}\cr
0 & i \Sigma_{1} & \epsilon_{fk}+ \Sigma_{0}-i\omega_{n} }
\right]\cr
&=& -\frac{T}{2}\sum_{\bk ,i\omega_{n}}\ln \left[
\epsilon_{fk}+\Sigma_{0}-i\omega_{n} 
 \right] 
-\frac{T}{2}\sum_{\bk ,i\omega_{n}}\ln \left[  (\epsilon_{fk}+\Sigma_{0}-i\omega_{n} )^{2}- \Sigma_{1}^{2}\right]
\end{eqnarray}
\end{widetext}
which has explicitly lost its dependence on the direction of $\bn$. We can thus be sure that even with a finite $\mu$ and
spin-orbit coupling the mean-field Free energy is isotropic. 

There will in general be a finite uniaxial anisotropy near the
surface, where broken inversion symmetry develops. 
However, provided the bulk maintains cubic isotropy, the 
isotropy of the Free energy required for a higher order-parameter
manifold is maintained at the mean-field level 
in the presence of spin-orbit coupling.

\end{document}